\documentclass[journal=jacsat,manuscript=article]{achemso}

\usepackage[version=3]{mhchem} 
\usepackage{hyperref}
\usepackage{lmodern}
\usepackage{amsfonts}

\newcommand{\ie}{i.e., }
\newcommand{\eg}{e.g., }

\usepackage[normalem]{ulem}

\newcommand{\res}{H_{\mathrm{res}}}
\newcommand{\rel}{H_{\mathrm{rel}}}
\newcommand{\nha}{n_{\mathrm{ha}}}
\newcommand{\ncg}{N_{\mathrm{CG}}}
\newcommand{\nopt}{N_{\mathrm{OPT}}}
\newcommand{\noptRM}{N_{\mathrm{OPT}}^{\mathrm{RM}}}
\newcommand{\noptIT}{N_{\mathrm{OPT}}^{\mathrm{IT}}}

\newcommand{\naa}{N_{\mathrm{AA}}}

\author{Margherita Mele}
\affiliation[Univerity]{Physics Department, University of Trento, via Sommarive, 14 I-38123 Trento, Italy}
\alsoaffiliation{INFN-TIFPA, Trento Institute for Fundamental Physics and Applications, I-38123 Trento, Italy}
\author{Raffaele Fiorentini}
\affiliation{Physics Department, University of Trento, via Sommarive, 14 I-38123 Trento, Italy}
\alsoaffiliation{INFN-TIFPA, Trento Institute for Fundamental Physics and Applications, I-38123 Trento, Italy}
\author{Thomas Tarenzi}
\affiliation{School of Chemistry, University of Birmingham, B15 2TT Birmingham, UK}
\author{Giovanni Mattiotti}
\affiliation{Physics Department, University of Trento, via Sommarive, 14 I-38123 Trento, Italy}
\alsoaffiliation{INFN-TIFPA, Trento Institute for Fundamental Physics and Applications, I-38123 Trento, Italy}
\alsoaffiliation{Unit\'e de Biologie Functionnelle et Adaptative, CNRS UMR 8251, Inserm ERL U1133, Universit\'e Paris Cit\'e, 35 rue H\'el\`ene Brion, Paris, France}
\author{Raffaello Potestio}
\affiliation{Physics Department, University of Trento, via Sommarive, 14 I-38123 Trento, Italy}
\alsoaffiliation{INFN-TIFPA, Trento Institute for Fundamental Physics and Applications, I-38123 Trento, Italy}
\email{raffaello.potestio@unitn.it}

\title
  {Determining the optimal structural resolution\\of proteins through an information-theoretic analysis of their conformational ensemble}


\begin{document}

\begin{tocentry}
\includegraphics[width=1\linewidth]{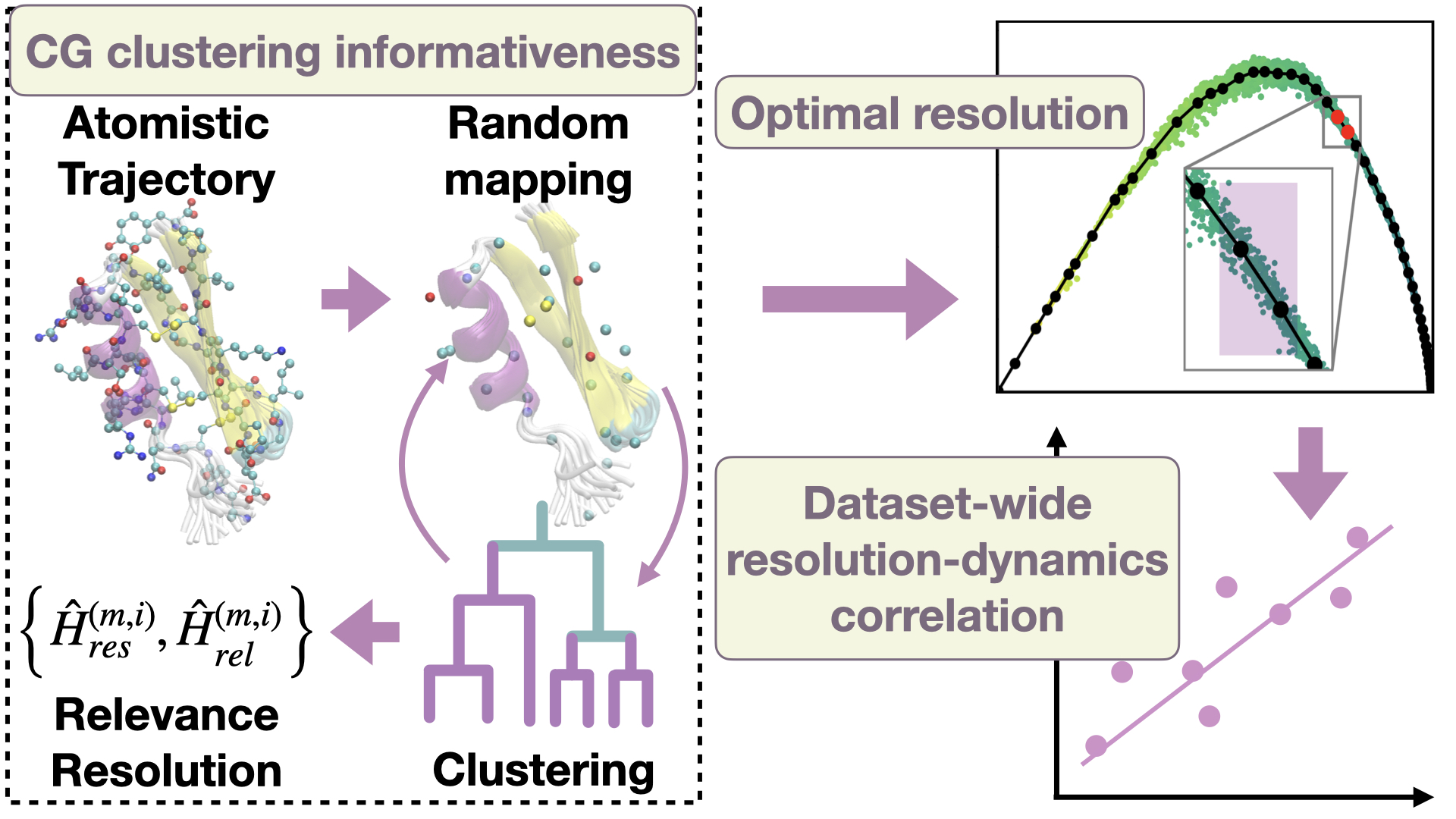}
\end{tocentry}
\begin{abstract}
The choice of structural resolution is a fundamental aspect of protein modelling, determining the balance between descriptive power and interpretability. Although atomistic simulations provide maximal detail, much of this information is redundant to understand the relevant large-scale motions and conformational states. Here, we introduce an unsupervised, information-theoretic framework that determines the minimal number of atoms required to retain a maximally informative description of the configurational space sampled by a protein. This framework quantifies the informativeness of coarse-grained representations obtained by systematically decimating atomic degrees of freedom and evaluating the resulting clustering of sampled conformations. Application to molecular dynamics trajectories of dynamically  diverse proteins shows that the optimal number of retained atoms scales linearly with system size, averaging about four heavy atoms per residue--remarkably consistent with the resolution of well-established coarse-grained models, such as MARTINI and SIRAH. Furthermore, the analysis shows that the optimal retained atoms number depends not only on molecular size but also on the extent of conformational exploration, decreasing for systems dominated by collective motions. The proposed method establishes a general criterion to identify the minimal structural detail that preserves the essential configurational information, thereby offering a new viewpoint on the structure-dynamics-function relationship in proteins and guiding the construction of parsimonious yet informative multiscale models.
\end{abstract}

\section{Introduction}
Proteins are complex biomolecules constituted by chains of amino acids, whose size ranges from a few tens to hundred thousands atoms \cite{ branden2012introduction, dill2012protein, labeit1995titins,leong2022short}. In order to carry out their biological function, a large number of proteins need a well-defined three dimensional structure, the native conformation, as well as a considerable degree of flexibility \cite{henzler2007dynamic,boehr2009role}. Hence, proteins populate their configurational space in a non-trivial manner, which is strictly intertwined with their function-oriented dynamics \cite{frauenfelder1991energy,dill1997levinthal, dill2008protein}.

Computer simulations \cite{mccammon1977dynamics,karplus2002molecular,dror2012biomolecular} represent a valuable, nowadays essential instrument to investigate protein structure, dynamics, and the interplay among them. Particularly, a key advantage of simulations over experimental approaches is the inherent high resolution of the molecule's representation, because in a computer model each atom can be explicitly accounted for. This allows one to track down the slightest conformational variation of the protein and its environment, and characterise the behaviour of the molecule down to its finest (classical) detail.

It has to be acknowledged, however, that for the sake of the analysis an extremely rich description of the system is not always necessary nor desirable: indeed, an excess of detail can obfuscate important large-scale features that are key to grasp the essential characteristics of the system. As an extreme example, the precise and exact knowledge of each atom of a protein is not useful in determining the relative population of two conformational basins; instead, fewer (even a single one) and much coarser variables might suffice \cite{best2005reaction,peters2006obtaining,noe2017collective}. This illustrates the important difference between data and information: one thing is to have many distinct \emph{configurations}, another thing is to group them together and see \emph{states} emerge \cite{chodera2014markov, noe2008transition}.

In computer simulations, this issue appears vigorously. Long, time-intensive molecular dynamics (MD) trajectories are necessary to achieve an adequately extensive coverage of the configurational space of a protein, and this need becomes especially painstaking as the size of the molecule increases. At the same time, though, the enormous number of distinct frames generated can be redundant with respect to certain information one might aim for. In a previous work \cite{mele2022information}, some of us have addressed this issue by grouping MD frames through a hierarchical clustering approach: varying the distance threshold below which two frames are grouped together, one can look for the level of configurational space coarsening that attains a compromise between data parsimony (smallest number of clusters) and informativeness (each cluster is representative of a distinct state). It was shown that an optimal trade-off point indeed exists, at which a large frame dataset can be compressed not only without losing important detail, but even revealing significant features.

The aforementioned work relied on a structure-based clustering of the protein conformations, in which either all heavy atoms or all C$_\alpha$ atoms were retained. Yet, the idea that the highest level of detail is not necessarily the most useful applies not only to the sampled state space, but rather also to the structure itself. In fact, while all atoms in the protein contribute to its emergent behaviour, not all of them contribute equally, because of the numerous internal constraints and restraints (chemical bonds, non-bonded interactions, free energy basins...) that limit the conformational variability of the degrees of freedom. Furthermore, the very existence and the successful application of coarse-grained models \cite{Noid2013_PerspectiveCG,saunders2013coarse, kmiecik2016coarse,giulini2021system} demonstrate that a simplified structural representation of proteins can suffice to reproduce, qualitatively and sometimes even quantitatively, many emergent properties that occur at a typical scale larger than that of the groups of atoms lumped into effective interaction sites. However, while it is relatively easy to embrace the idea that a low-resolution representation of a protein can be useful and informative, \emph{finding} such a representation is a much more complex endeavour. The question thus arises, whether it is possible to quantify in an unambiguous manner the informativeness of a coarse-grained representation of a protein and, by maximising it, to identify an \emph{optimal resolution level}, that is, the optimal number of atoms to be preserved in the low-resolution representation.

In this work, we take the moves from the idea that a valid, useful coarse-grained description of a protein is one that allows a statistically significant clustering of its configurational space. Such clustering is obtained grouping together high-resolution configurations that map onto the same low-resolution representation: if the representation is too detailed, the dataset will be partitioned in a large number of small, statistically uninformative clusters; in contrast, a too coarse description will merge frames that should be ascribed to different states, thereby losing the capability to discriminate physically meaningful structural differences.

In practice, we here address this problem as follows. First, as coarse-grained representations we consider \emph{decimation mappings} \cite{giulini2020information,ChaimovichShell2011_RelEntMapping} of a protein, that is, selections of a subset of its heavy atoms, and we investigate the properties of the molecule in terms of the number of such atoms. Second, given an MD simulation of the protein, we cluster the sampled conformations based on the selected subset, and repeat the clustering for many different atoms at the same resolution level. Lastly, we quantify the \emph{informativeness} of each clustering and search for the resolution level that, on average, maximises the retained information while minimising the detail.

Key to this last step is evidently a quantitative notion of informativeness. There is no unique way in which one can measure the amount of information entailed by a dataset, and the most adequate approach is generally problem-specific. Here, it is our objective to establish how informative a given clustering is about the generative process that underlies the sampled MD frames, implying, by this, that the statistical weights of the clusters reverberate the depths of the associated free energy basins. To do so, we employ critical variable selection \cite{grigolon2016identifying, marsili2022quantifying}, an information theory approach that quantifies, with distinct metrics, the level of detail of the representation of a given ensemble (its resolution) and the amount of useful information it contains about the stochastic process that generated the data (the relevance).

Based on these tools we have constructed an unsupervised procedure, dubbed protein optimal resolution identification method, or PROPRE for short, that processes an MD trajectory of a protein (but in principle is applicable to any macromolecule) and returns the optimal number of atoms that one should retain to build a synthetic yet maximally informative low-resolution representation of the system. The approach was applied to a dataset of structurally and dynamically different globular proteins, as well as to a protein that interconverts, in the course of the same simulation, between two conformational states (open/closed). The analysis we carried out showed that the optimal number of atoms to retain for a maximally informative reduced representation of a protein scales linearly with its size, but it also has a nontrivial relation with the conformational properties of the molecule. In fact, this number varies depending on the specific conformational basin that the protein populates. Lastly, the optimal number of retained atoms identified through the PROPRE approach marks the resolution level at which a targeted optimization can return the most informative coarse-grained representation.

The paper is organized as follows. First, we illustrate the theoretical foundations of the PROPRE algorithm and its technical workflow; subsequently, we showcase its application to nontrivial case studies; finally, we discuss its results. The method is implemented in an agile and user-friendly python script that is made freely available for download and usage in the repository \href{https://github.com/potestiolab/propre}{https://github.com/potestiolab/propre}.
\begin{figure}[H]
    \centering
    \includegraphics[width=0.5\textwidth]{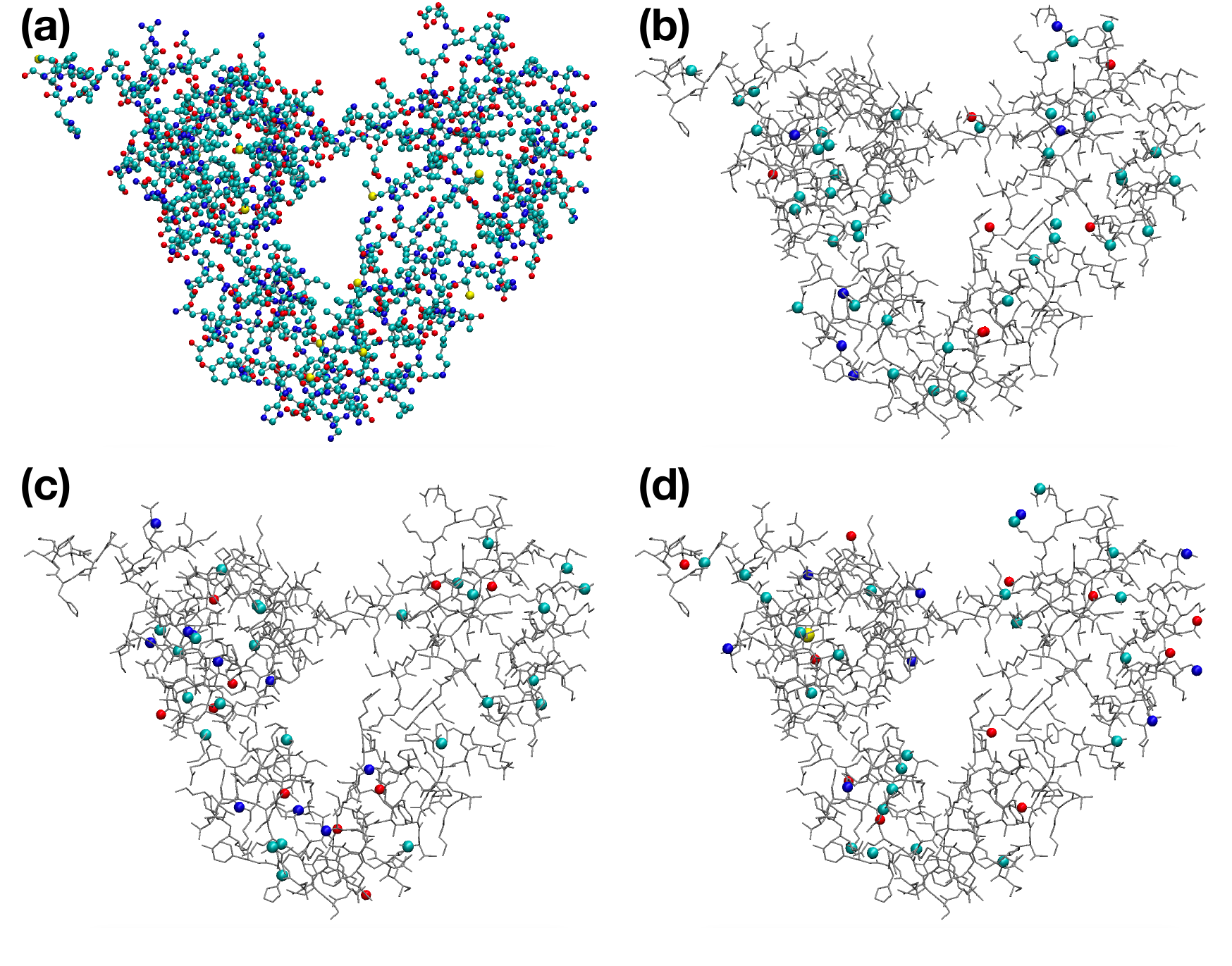}
    \caption{Example of different decimation mappings of a protein. Panel \textbf{(a)} shows a protein with $N_{ha} = 2027$ heavy atoms. In panels \textbf{(b)}, \textbf{(c)}, and \textbf{(d)} $3$ different decimation mappings of the same protein are shown, selecting a subset of $N_{CG} = 100$ atoms out of $N_{ha}$.}
    \label{fig:mappings}
\end{figure}

\section{The PROPRE method: theoretical background and algorithmic aspects}

The PROPRE approach consists in a protocol for the analysis of an atomistic trajectory of a biomolecule, aimed at identifying the optimal number $\nopt$ of heavy atoms to retain in a synthetic yet informative low-resolution description of the system and its conformational space. In other words, the method allows the determination of the \textit{optimal level of resolution} to capture the essential features of the biomolecule's large-scale dynamics, retaining the smallest amount of detail while preserving the largest amount of information.

In the following sections, the notions of \emph{amount of detail} and \emph{amount of information} will become clearer and quantitative, and we will describe the theoretical foundations of PROPRE and its workflow. The software implementing the method, including the programme itself and examples of input files, are freely available in the repository \\
\href{https://github.com/potestiolab/propre}{https://github.com/potestiolab/propre}. The repository includes three Python scripts: the first one removes all hydrogen atoms from both the molecule configuration file and the trajectory file; the second script computes the level of informativeness of various low-resolution representations as a function of the number of retained atoms, making use of two quantities, resolution and relevance, which are discussed in detail in the next subsection; the third script employs the results of the second to pinpoint the optimal number of sites; lastly, the script outputs the processed data and includes example code to reproduce the plots. A more detailed description of script usage and several examples of input files are available as \texttt{readme} files in the repository itself.
\subsection{Critical variable selection, or resolution and relevance framework}

The PROPRE protocol builds on the \emph{critical variable selection} (CVS), or \emph{resolution-relevance} framework, introduced by Marsili and co-workers \cite{marsili2013sampling, cubero2019statistical, marsili2022quantifying} and applied across diverse domains, from protein sequences \cite{grigolon2016identifying} to neural activity \cite{cubero2020multiscale} and protein conformational ensembles \cite{mele2022information}. This approach provides an information-theoretic criterion to identify informative coarse representations of complex datasets. Detailed theoretical discussions can be found elsewhere \cite{marsili2013sampling, cubero2019statistical, marsili2022quantifying, mele2022information, holtzman2022making}; here we summarize the key definitions and their role within PROPRE workflow.

Consider a dataset of $M$ elements, such as MD frames, partitioned into $L$ clusters labeled by $s$. The \emph{resolution} of this partition is defined as the entropy of cluster occupancies,
\begin{equation}
\res[s] = -\sum_{s=1}^L p_s \log p_s, \qquad p_s = \frac{k_s}{M},
\label{eq:resolution}
\end{equation}
where $k_s$ is the number of elements in cluster $s$. Resolution quantifies the level of detail, ranging from $\res=0$ when all elements belong to a single group to $\res=\log M$ when every element is distinct.
The complementary notion of \emph{relevance} captures how informative the clustering is, by measuring the heterogeneity in cluster sizes \cite{marsili2013sampling,marsili2022quantifying}:

\begin{equation}
\rel[k] = - \sum_{k=1}^M \frac{k\, m_k}{M} \log \left(\frac{k\,m_k}{M}\right),
\label{eq:relevance}
\end{equation}
with $m_k$ the number of clusters containing $k$ elements. Intuitively, while resolution reflects how finely the dataset is partitioned, relevance probes whether the partition reveals meaningful structure: heterogeneous cluster occupancies correspond to non-uniform sampling of configuration space, and therefore to underlying modulation of the probability distribution shaped by the system's free energy landscape; in contrast, a partition of the sample in similarly-sized clusters does not suggest features of sort. A high relevance thus indicates that the data explores distinct regions with different statistical weights, while a low relevance corresponds to partitions that obscure such structure.

Relevance vanishes at both extremes of resolution, since both the trivial partition (all elements in one cluster) and the maximally detailed one (all singletons) evidence no statistical structure. Between these extremes, $\rel[k]$ attains a maximum, which identifies the clustering that retains the largest amount of non-redundant information about the data. The relation between resolution and relevance takes the form of a bell-shaped curve, whose interpretation is central to the framework. The maximum of the curve corresponds to the most informative compression of the dataset, where heterogeneity in cluster occupancies is most pronounced ($\nopt^{\mathrm{MR}}$). Equally important is the point where the slope of the curve equals $\mu = -1$ ($\nopt^{\mathrm{IT}}$), which signals an optimal trade-off: here a unit loss in resolution is exactly balanced by a unit gain in informativeness \cite{marsili2013sampling,marsili2022quantifying}. From an information-theoretic perspective, these two points highlight the most meaningful coarse-grained descriptions of the dataset, either by maximizing relevance or by ensuring an optimal balance between detail loss and information gain.

Within PROPRE, this framework is employed to evaluate the extent to which coarse representations of a protein's structure preserve information about its sampled conformational space. By focusing on the maximum relevance and the $-1$ slope points, one can identify atom subsets and, consequently, clustering levels that yield the most informative yet parsimonious description of the protein.
\subsection{Calculation of the resolution-relevance curves}
\begin{figure*}
    \centering
    \includegraphics[width=0.9\textwidth]{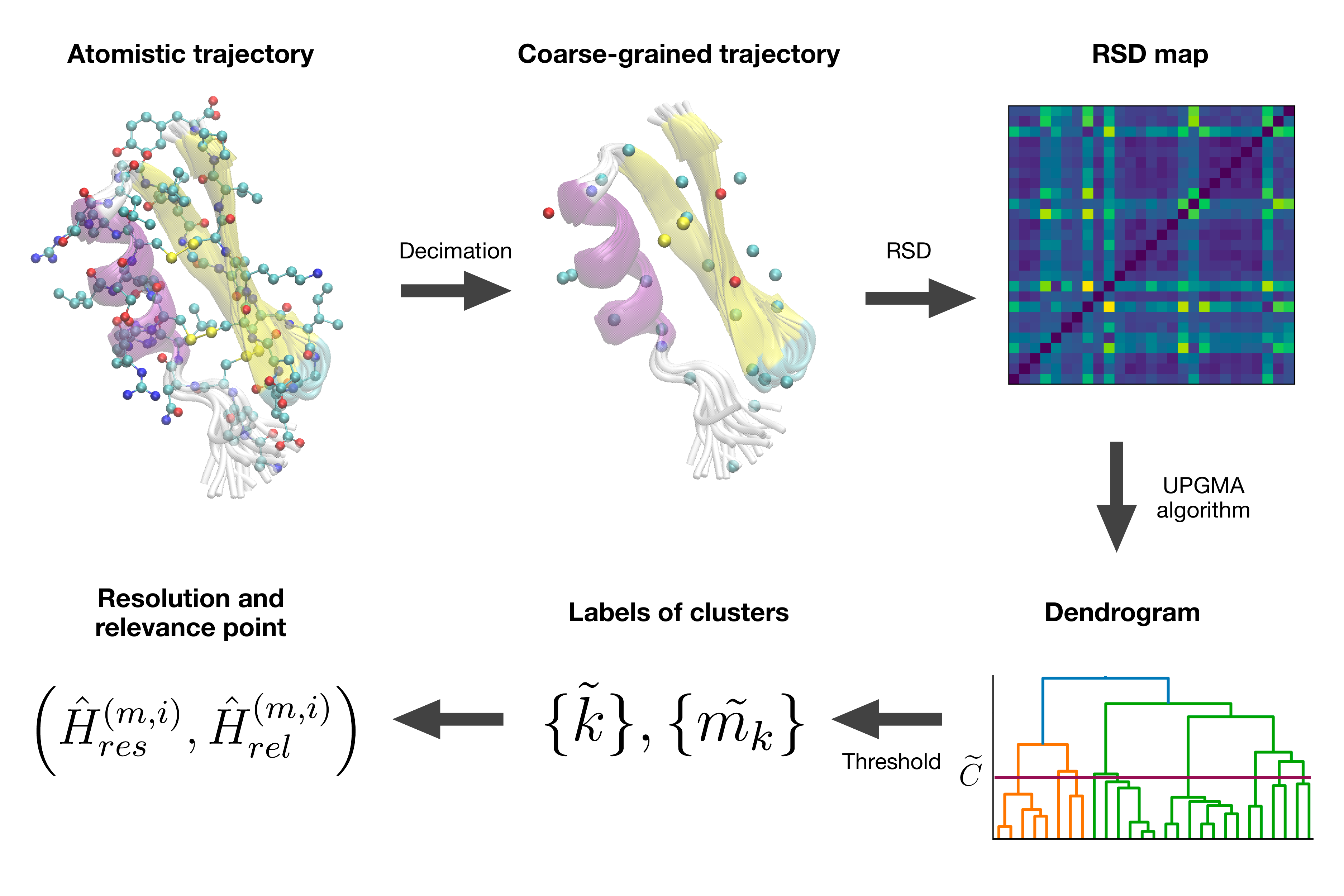}
    \caption{Infographic that illustrates the steps of the first part of the PROPRE protocol: the construction of the $\res,\rel$ scatter plot. After filtering the atomistic trajectory (by choosing a random decimation mapping of the heavy atoms), the RSD map is built, in order to perform a clustering of the filtered configurations. Each mapping corresponds to a point in the $\res,\rel$ plane; by varying the retained atoms selection as well as the number of retained atoms one can reconstruct the curve as a scatter plot.}
    \label{fig:infographic}
\end{figure*}

The algorithm scans a range of numbers of retained (heavy) atoms, denoted $ \ncg$, and computes, for each value, the corresponding resolution and relevance over a sample of randomly generated mappings. A schematic illustration of the whole workflow is provided in Fig.~\ref{fig:infographic}. 

As anticipated, the starting point of the PROPRE algorithm is the molecular dynamics trajectory of a biomolecule, which contains all atoms of the system of interest but not those of the solvent. Denoting with $\nha$ the number of heavy atoms in the molecule, the atomistic trajectory $\mathbf{X}_\mathrm{ha}(t)$ consists of a collection of $T$ vectors of dimension $3\nha$, each labelled with the time index $t$ associated with the given time frame. In the applications presented here, we use only the heavy-atom coordinates; other non-system-specific features could be incorporated without affecting its generality or unsupervised character.

The UPGMA \cite{sokal1958statistical} hierarchical clustering algorithm is then applied to cluster the configurations of $\mathbf{X}_\mathrm{ha}(t)$. The metric employed to measure the distance between two configurations $\mathbf{X}^a$ and $\mathbf{X}^b$ is the root squared distance (RSD), defined as:
\begin{eqnarray}
&& \mathrm{RSD} = \sqrt{ \sum_{i = 1}^{\nha} |\mathbf{X}_i^a - \mathbf{X}_i^b|^2}.
\end{eqnarray}

The RSD is proportional to the more common root mean squared distance, $\mathrm{RSD} = \sqrt{\nha}\cdot \mathrm{RMSD}$, which, despite known limitations \cite{shao2007clustering,maiorov1994significance}, still represents the most reliable, interpretable, and commonly used measure to quantify the dissimilarity between two protein structures \cite{kufareva2012methods}.

The choice of employing the RSD is motivated by advantages it entails in the present context \cite{giulini2020information,holtzman2022making}. Specifically, if we employed the RMSD as a measure of distance between pairs of configurations, we would have to set, for each value of $\ncg$, a different clustering threshold, since the RMSD is normalized by the number of sites. By contrast, the usage of the RSD preserves the absolute scale of distances, allowing us to choose a single threshold that can be employed irrespective of the number of sites retained at each iteration step. The aforementioned RSD distance threshold $\Tilde{C}$ is the minimal value of dissimilarity, that is, the smallest distance below which two configurations can be considered to be structurally identical. To set this value, we perform a hierarchical clustering of the all-atom trajectory, and define $\Tilde{C}$ as the largest value below which all frames are distinguishable, i.e. each frame is representative of a singleton cluster. The threshold $\Tilde{C}$ is then employed in the next step of the workflow; here we consider a number $\ncg^{(i)} = \nha - i$ of retained atoms for $i \in [2, \nha]$. A number $\mathcal N_M$ of different \emph{mappings} is generated, that is, a random selection of $\ncg^{(i)}$ heavy atoms out of the $\nha$ available ones. For each one of the $\mathcal N_M$ mappings, the trajectory $\mathbf{X}_\mathrm{ha}(t)$ is processed to obtain the CG configuration $\mathbf{X}^{(m,i)}_\mathrm{CG}(t)$ of the sole atoms that are included in the mapping, the index $m$ labelling the specific mapping and $i$ indicating the retained atom number $\ncg^{(i)}$. The frames of this trajectory are clustered with the UPGMA algorithm, and the dendrogram is cut at the threshold value $\Tilde{C}$: in this way, two CG configurations are considered indistinguishable if their distance falls below the same threshold used to resolve structural differences in the all-atom space. This ensures a consistent resolution criterion across both representations: CG configurations are grouped together only if the retained structural differences would have been negligible at full resolution, and treated as distinct only if their full-atom counterparts were also distinguishable.
\begin{figure}[H]
    \centering
    \includegraphics[width=0.5\textwidth]{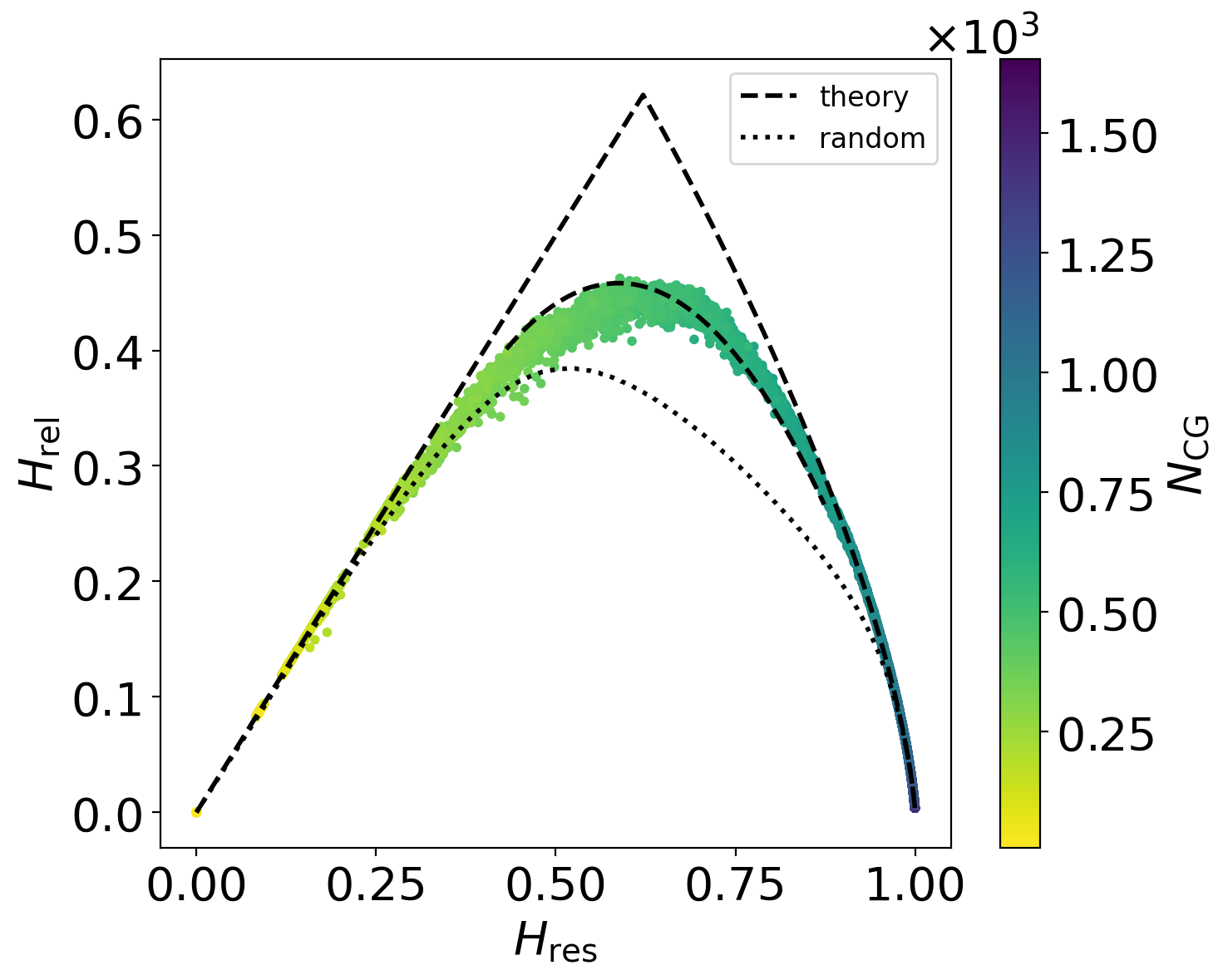}
    \caption{Plot of resolution ($\res$) vs. relevance ($\rel$) obtained for a trajectory of the enzyme adenylate kinase. Points are coloured according to the number of atoms retained in the corresponding mapping $\ncg$. The black dashed lines represent the upper and lower bound to the theoretical maximum, derived in \cite{marsili2013sampling, haimovici2015criticality}. The grey dotted line shows the typical behaviour of a structure-less sample obtained by averaging over multiple random partitions of M balls in an increasing number of boxes.} 
    \label{fig:HsHkAKE}
\end{figure}

The clusters so obtained are then used to calculate one $(\res,\rel)$ point; see Eqs. \ref{eq:resolution} and \ref{eq:relevance}. This procedure is repeated for each choice of the number of retained atoms $\ncg^{(i)}$ and for each random mapping $m$ generated retaining $\ncg^{(i)}$ atoms. The resulting points draw a curve in the resolution-relevance plane as the one shown in Fig.~\ref{fig:HsHkAKE} for the protein \textit{adenylate kinase} (see below). These data can then be analysed to identify ``special'' resolution values, namely the point where the curve has slope $-1$ ($\nopt^{\mathrm{IT}}$) and the point corresponding to the maximum of the relevance ($\nopt^{\mathrm{MR}}$). In the following, we will present the results obtained employing both criteria unless otherwise stated; the algorithmic procedure used to identify such values, including smoothing and numerical differentiation of the resolution-relevance curve, is described in the Supplementary Information (SI).

\section{Materials and methods}
\subsection{Protein selection}
A dataset of 11 proteins distinct in size, structure and function was built starting from a larger dataset of 107 proteins, after clusterization based on their dynamics. The initial, large dataset was based on the one employed in reference \cite{hensen2012exploring}, from which proteins with too high sequence/structure similarity were removed. The proteins are categorized monomeric by the protein quaternary structure file server (PQS), have a resolution higher than 1.8\AA, no ligands larger than six atoms, and no presence of metals other than Mg$^{2+}$, Ca$^{2+}$, K$^{+}$, Na$^{+}$, Zn$^{+}$. Further requirements included a structure deposition date after 1987 and absence of gaps larger than one amino acid. The structures were downloaded from the Protein Data Bank, and the coordinate files were cleaned-up from heteroatoms, from copies of the protein in the crystallographic cell, and from residue-configurations with low occupancy. The position of missing atoms was rebuilt and the protein conformations were optimized using the software FoldX \cite{schymkowitz2005foldx}. The procedure of dynamics-based clustering applied on this dataset reflects the one developed by us in \cite{tarenzi2022search}, and already applied in \cite{mele2022information}. Specifically, for each protein of the larger dataset, the first 10 normal modes of fluctuation were analysed using an elastic network model, and superimposed by means of the ALADYN protocol \cite{potestio2010aladyn}, which performs a hybrid structural/dynamical alignment. The similarity between the essential spaces spanned by the first 10 normal modes was quantified by means of the root mean square inner product (RMSIP) \cite{david2011characterizing}. The distance between the essential dynamics of two aligned proteins was defined as $d_{ij} = 1 - \textrm{RMSIP}_{ij}$; this distance was employed to perform a hierarchical clustering. The 11 proteins used in this work are the centroids of the resulting clusters, and their PDB codes are: 1DSL, 1NOA, 1SNO, 1UNE, 1XWL, 1IGD, 1HYP, 1KNT, 1QKE, 2EXO, 1KOE.

In addition to this dataset, we also employed the enzyme adenylate kinase (ADK) to study the dependence of the optimal number of retained atoms on the protein conformational state. ADK was chosen because of its large conformational variability, which can be sampled on timescales accessible to standard atomistic MD simulations. The starting structure for the MD simulations corresponds to the crystallographic structure of ADK in the open form (PDB ID: 4AKE).

\subsection{Simulation and analysis protocols}
All simulations were performed with the Gromacs 2018 software  \cite{abraham2015gromacs}. The protein topology was defined through the AMBER99SB-ILDN force field \cite{lindorff2010improved}, and the TIP3P model was employed for the description of water molecules \cite{jorgensen1983comparison}. Sodium and chloride ions were added at the physiological concentration of $150$ mM, neutralising the global electric charge in the simulation box. After energy minimisation, NVT and NPT equilibrations were performed using the velocity-rescale thermostat \cite{bussi2007canonical} and the Parrinello-Rahman barostat \cite{parrinello1981polymorphic}. A cut-off of $1.2$ nm was used for van der Waals interaction and for the short-range component of the Coulomb one. The long-range component of the Coulomb force, instead, was computed with the Particle Mesh Ewald algorithm. The LINCS \cite{hess1997lincs} algorithm was employed to define the constraints on the hydrogen-containing bonds. An integration time step of $2$ fs was used. For each protein of the dataset, a trajectory of $500$ ns was collected; of these, the first $100$ ns are considered as equilibration and therefore discarded.

Analysis of the trajectories was performed with Gromacs tools and in-house scripts. Protein residue fluctuations were calculated with \textit{gmx rmsf}, while the clustering and the identification of the representative conformations was performed with \textit{gmx cluster}. The rendering of protein structures was performed with VMD \cite{humphrey14vmd} and with TCL scripts.

\begin{figure*}
\centering
\includegraphics[width=0.9\textwidth]{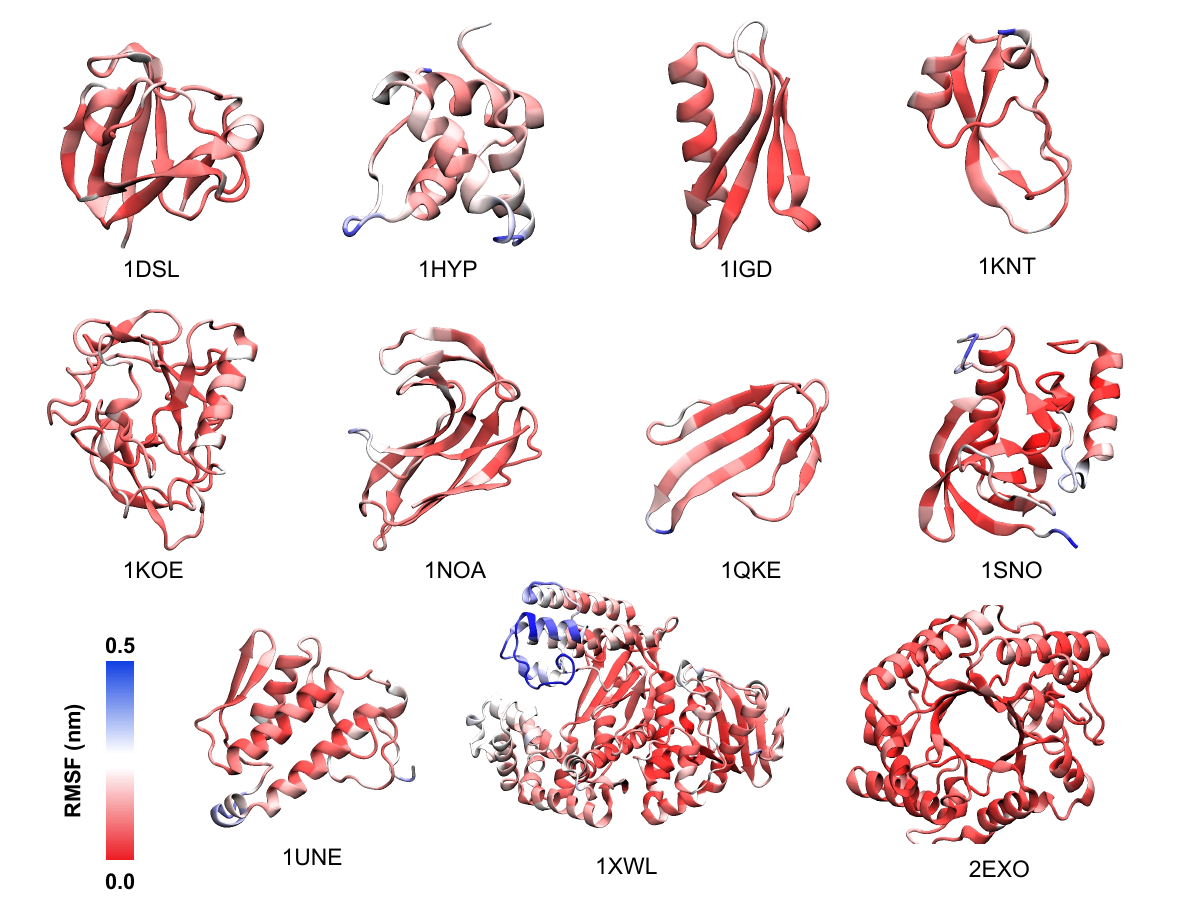}
\caption{Graphical representation of the proteins comprising the dataset, colored according to the per-residue value of root-mean-square fluctuations as computed from the MD simulations.}
\label{fig:dataset_rendering}
\end{figure*}

\subsection{PROPRE parameters}

For each simulation, PROPRE analysis is performed on $1000$ frames, selected either as equally spaced along the trajectory (uniform sampling) or as centroids from UPGMA clustering (centroid sampling). In accordance with the default software parameters, the number of random mappings for a fixed number of CG sites is $50$, while the number of retained atoms itself is progressively decreased with a step corresponding to $0.5$\% of the total number of protein heavy atoms. From the computed relevance-resolution curves, a smoothing procedure is applied to identify the points of largest relevance and slope $-1$; the detailed procedure and the associated relevant parameters are illustrated in SI, Sec.II.

\subsection{Calculation of the covariance matrix}
\label{sec:met_cov}
To quantify the structural fluctuations of a molecular system composed of $\nha$ heavy atoms over $T$ time steps, we compute the covariance matrix of atomic positions. Let \( \mathbf{P} \in \mathbb{R}^{T \times 3\nha} \) denote the matrix of atomic configurations, where each row corresponds to the 3D positions of all heavy atoms at a given time step, flattened into a vector of length $3\nha$. The mean configuration over time is given by \( \bar{\mathbf{P}} \in \mathbb{R}^{1 \times 3 \nha} \). The covariance matrix \( \mathbf{C} \in \mathbb{R}^{3 \nha \times 3 \nha} \) is then defined as:

\[
\mathbf{C} = \frac{1}{N - 1} (\mathbf{P} - \bar{\mathbf{P}})^\top (\mathbf{P} - \bar{\mathbf{P}}).
\]

This matrix captures the variance and covariance of positional fluctuations, including both intra- and inter-atomic correlations among Cartesian coordinates.
As a scalar descriptor of the overall fluctuation magnitude, we consider the \emph{trace} of the covariance matrix, defined as the sum of its diagonal elements:
\[ \mathcal{T}^{\mathrm{ha}}
 = \mathrm{Tr}(\mathbf{C}) = \frac{1}{T - 1} \sum_{i=1}^{T} \left\| \mathbf{P}_i - \bar{\mathbf{P}} \right\|^2.
\]

The same construction can be applied when only $\ncg$ atoms are kept, chosen at random among the $\nha$ heavy atoms. The corresponding fluctuation measure is denoted $\mathcal{T}^{\mathrm{CG}}$. 
Since the trace grows proportionally with the number of atoms, the two quantities $\mathcal{T}^{\mathrm{ha}}$ and $\mathcal{T}^{\mathrm{CG}}$ are not directly comparable. We therefore introduce a size-independent observable
\[
\mathrm{cov} = \frac{\mathcal{T}^{\mathrm{CG}} \cdot \nha }{\mathcal{T}^{\mathrm{ha}} \cdot \ncg}.
\]

This normalisation accounts for the difference in system size by assigning an average fluctuation contribution to the atoms excluded during coarse-graining. By construction, \( \mathrm{cov} = 1 \) when all atoms are retained, and values near 1 indicate that the retained subset captures the overall structural variability of the whole system. The behaviour of \( \mathrm{cov} \) as a function of \( \ncg \) provides a diagnostic for quantifying how representative an atoms subset is of the system's fluctuations.

\subsection{Mapping entropy}
\label{sec:mapping_entropy}

Mapping entropy quantifies the intrinsic loss of information that arises when a fine-grained system is projected onto a CG representation. This concept was first introduced within the framework of the relative entropy method \cite{shell2008relative} and has since been widely applied to the analysis of proteins as well as other systems \cite{giulini2020information,holtzman2022making,giulini2021system,aldrigo2025low}.

Let \( \mathbf{x} \in \mathbb{R}^{3\nha} \)  denote a microscopic configuration of a molecular system composed by \(\nha\) heavy atoms, sampled from the canonical distribution \( p(\mathbf{x}) \propto e^{-\beta U(\mathbf{x})} \), where  \( U(\mathbf{x}) \) is the potential energy and \( \beta = 1/(k_B T) \) the inverse temperature. A CG representation can be defined by selecting a subset of \( \ncg < \nha \) atoms. The selection is encoded by a binary vector \( \boldsymbol{\sigma} = \{ \sigma_i \in \{0,1\} \}_{i=1}^{\nha} \) such that \( \sum_{i=1}^{\nha} \sigma_i = \ncg \). This defines a projection operator \( M \) that maps the full configuration \( \mathbf{x} \) to a coarse-grained macrostate \( \mathbf{X} = M(\mathbf{x}) \), retaining only the positions of the selected atoms.

The mapping entropy \( S_{\mathrm{map}} \) is defined as the Kullback-Leibler divergence between the original microscopic distribution \( p(\mathbf{x}) \) and the reconstructed distribution \( \tilde{p}(\mathbf{x}) \) inferred from the coarse-grained representation:

\[
S_{\mathrm{map}} = k_b \int p(\mathbf{x}) \ln \left( \frac{p(\mathbf{x})}{\tilde{p}(\mathbf{x})} \right) \mathrm{d}\mathbf{x}.
\]

The reconstructed distribution \( \tilde{p}(\mathbf{x}) \) corresponds to the statistical model obtained by backmapping from the CG variables and is given by:

\[
\tilde{p}(\mathbf{x}) = \frac{p_{\mathrm{CG}}(M(\mathbf{x}))}{\Omega(M(\mathbf{x}))},
\]
where \( p_{\mathrm{CG}}(\mathbf{X}) = \int p(\mathbf{x}') \delta(M(\mathbf{x}') - \mathbf{X}) \, \mathrm{d}\mathbf{x}' \) is the induced probability over coarse-grained configurations, and \( \Omega(\mathbf{X}) = \int \delta(M(\mathbf{x}') - \mathbf{X}) \, \mathrm{d}\mathbf{x}' \) is the degeneracy of the macrostate \( \mathbf{X} \), i.e., the volume of microscopic configurations such that \( M(\mathbf{x}) =  \mathbf{X} \).

This functional is non-negative and vanishes only when the CG representation preserves the full statistical structure of the original representation. In this work, we evaluate \( S_{\mathrm{map}} \) for a sample of randomly generated selections of \( \ncg \) atoms, drawn from the \( \nha \) available atoms, to estimate the average information loss associated with unsupervised, non-optimized coarse-graining choices.

Mapping entropy calculations were performed using the EXCOGITO software package \cite{giulini2024excogito}, a general-purpose toolkit for coarse-graining analysis. In our case, the mapping entropy was estimated as the Kullback-Leibler divergence between the full microscopic distribution and the reconstructed one, with structural similarity assessed through clustering based on the root square deviation (RSD), adopting the same distance threshold as in the PROPRE protocol. Further details on the methodology can be found in \cite{giulini2024excogito, holtzman2022making, shell2008relative, chaimovich2011coarse}.


\section{Results and discussion}

\subsection{Optimal number of sites \textit{vs} number of protein residues}\label{Nopt-Nres}

The PROPRE algorithm allows the identification of the optimal resolution for a specific protein, where the optimal resolution is intended as the fraction of the total number of sites (namely, of the protein heavy atoms) that has to be retained in order to minimize the information loss upon coarse-graining. In the following, the optimal number of retained sites is the one that corresponds to slope $-1$ in the resolution-relevance curve; analogous results obtained maximising the relevance are reported in the SI.

\begin{figure*}
\centering
\includegraphics[width=\textwidth]{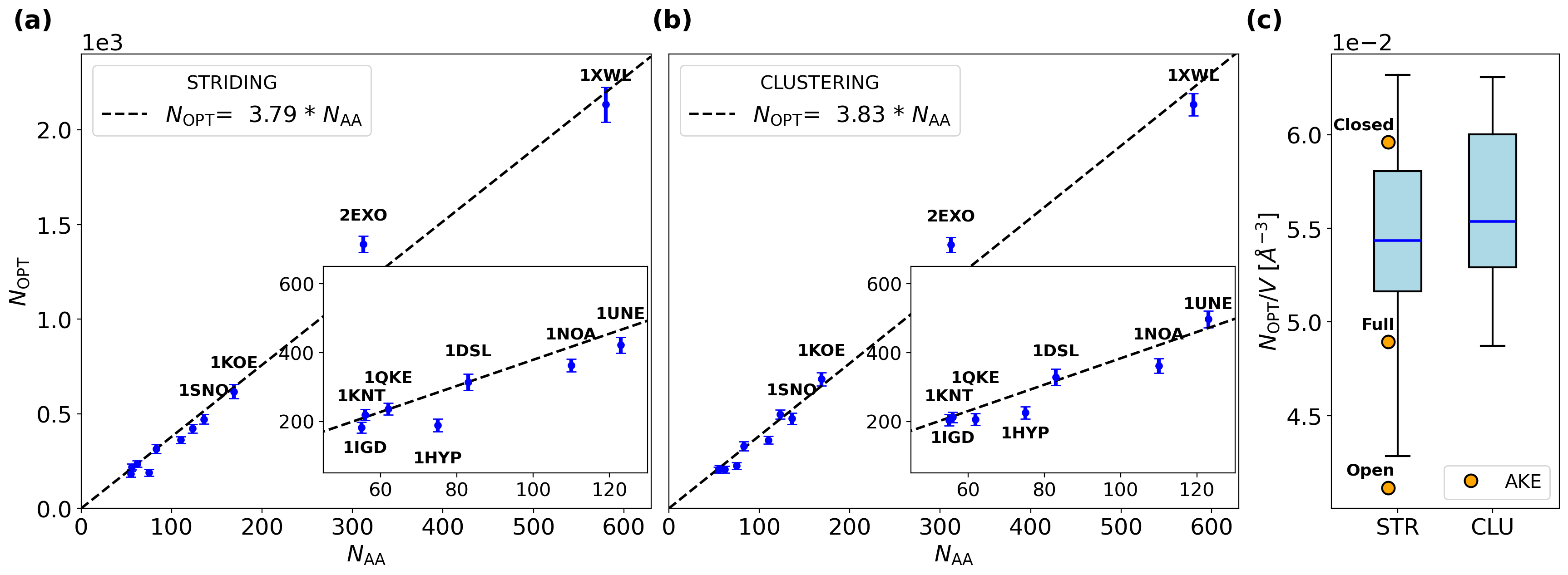}
\caption{(a,b) Scatter plots illustrating the number of optimal sites identified by PROPRE for each protein in relation to the number of residues; for both cases (striding and clustering) the correlation coefficient of the linear fit ($R^2$) is $0.98$, the Pearson correlation coefficient is $0.99$. (a) Analysis based on $1000$ equidistant frames extracted through striding. (b) Analysis based on the centroids of $1000$ clusters obtained from a UPGMA clustering procedure. Error bars indicate the standard deviation in the number of optimal sites within the most probable bin, as determined by PROPRE using the density-based protocol. (c) Box plots of the ratio between the number of optimal sites and the average sampled volume ($\nopt / V$)  for the two strategies: striding (STR) and clustering (CLU). For STR, orange points indicate values for the AKE protein computed from the full trajectory, as well as separately for the open and closed conformations.}
\label{fig:dataset_nres}
\end{figure*}

It is natural to think that the optimal number of sites $\nopt$ depends on the specific system under investigation and its features, e.g. the protein size, function, and flexibility. To assess the possible presence of a general trend, we fist applied the PROPRE pipeline to $500$ ns-long MD trajectories of a set of $11$ different proteins (see Fig.~\ref{fig:dataset_rendering}), ranging from a molecular weight of $7$ kDa (1IGD) to $66$ kDa (1XWL). The systems were selected from a larger protein dataset in order to maximize the differences in tertiary structure and dynamical features (see Methods section). Despite the large differences in structure, size and dynamics, the analysis on the trajectories of the 11 proteins of the dataset suggests a direct link between the optimal number of retained sites ($\nopt$) and the total number of protein residues ($\naa$).  Fig~\ref{fig:dataset_nres} (a,b) reports this relationship for two distinct strategies to sample the trajectory: striding, that is, retaining $1000$ equally spaced frames; and clustering, i.e. employing the $1000$ centroids of an UPGMA clustering.  In both cases, a clear linear trend emerges, with Pearson correlation coefficients $R = 0.99$ for both uniform and centroid sampling. The fitted slope of about $3.8$ in both cases reveals a systematic trend, whereby optimal coarse-graining retains almost four heavy atoms per residue, suggesting that a description of backbone plus minimal side chain captures the essential degrees of freedom.

While system size is a key determinant of the optimal resolution, it is not the only contributing factor: as it will be discussed later, it also depends on intrinsic dynamical features, such as the extent of conformational variability and the diversity of the explored structural states. In the present dataset, none of the proteins undergoes major conformational changes, which likely explains the consistency between the two sampling protocols. Differences are expected to emerge when analysing trajectory segments associated with distinct conformational states, as the optimal resolution may vary across structurally heterogeneous regions of configurational space. 

A complementary way of looking at the same trend is to normalise the optimal number of retained sites by the average configurational volume explored during the simulation ($V$). The resulting values of $\nopt / V$, reported in Fig.~\ref{fig:dataset_nres}.c, are narrowly distributed among the 11 proteins, despite their differences in size and topology. This points to the presence of a characteristic density, suggesting that proteins with stable globular folds tend to retain a consistent number of optimal sites per unit of sampled volume.

\begin{figure}
    \centering
    \includegraphics[width=0.5\linewidth]{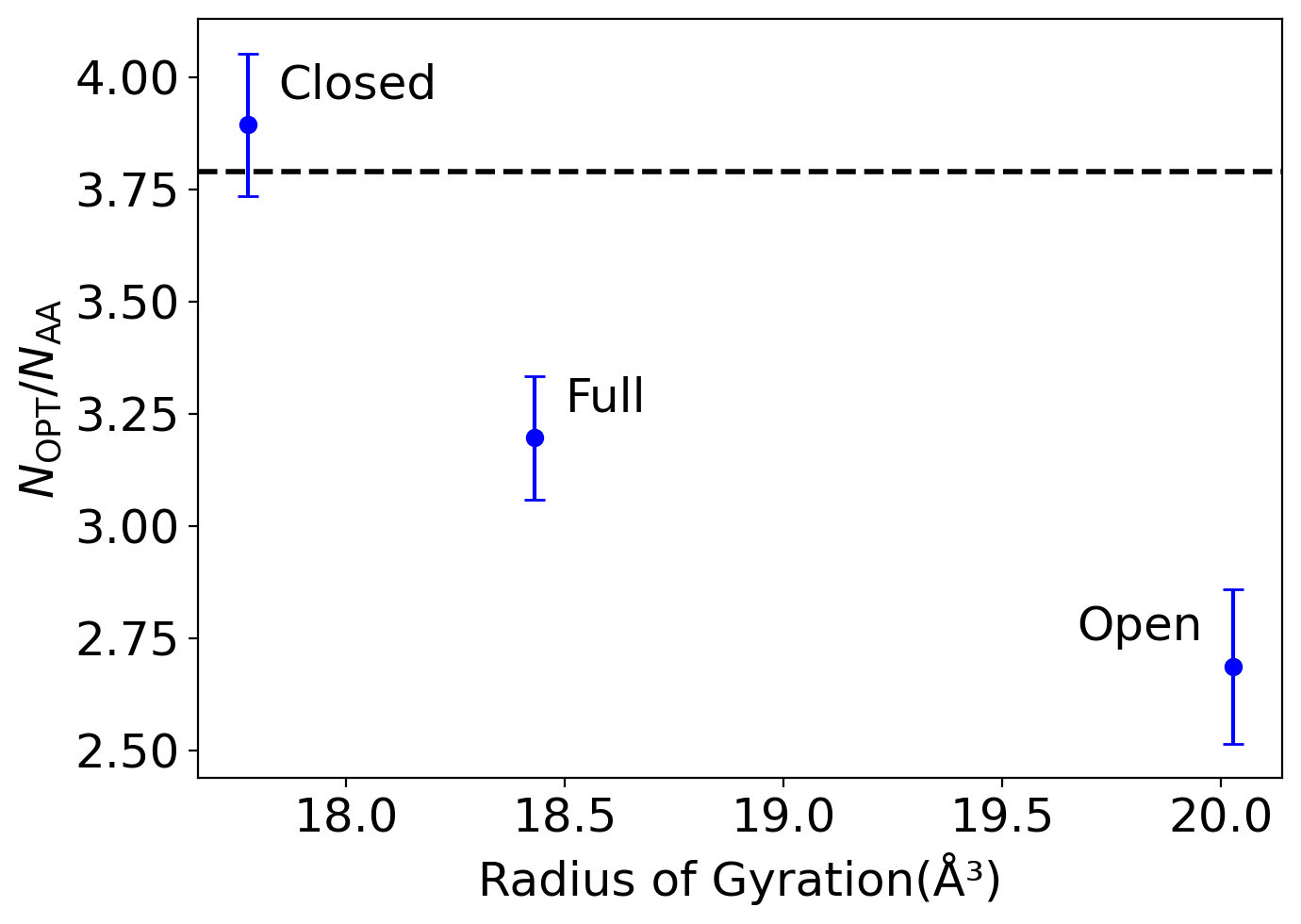}
    \caption{Scatter plot of the ratio between the optimal site number and residue number ($\nopt / \naa$) identified for adenylate kinase from the full trajectory and from the subsets corresponding to the open and closed states, as a function of the average radius of gyration sampled in each case. The optimal atom number corresponds to the resolution point of slope $-1$. The dashed horizontal line indicates the reference value fitted on the 11-protein dataset.}
    \label{fig:ake_volume}
\end{figure}

To explicitly probe the role of conformational variability, we turned to adenylate kinase (AKE), a paradigmatic system undergoing large-scale transitions between open and closed states. The MD trajectory was analysed in three different ways: the full trajectory, the open-conformation substate, and the closed-conformation substate, the latter two obtained \emph{via} UPGMA clustering. Despite the system being the same in all cases, each subset yields a distinct optimal number of retained atoms, $\nopt$.  In the box plot of Fig.~\ref{fig:dataset_nres}c, all three AKE cases fall within the same density range as the $11$ proteins, since variations in sampled volume are accompanied by corresponding changes in $\nopt$. 

A different picture emerges when considering the ratio $\nopt/\naa$ as a function of the average radius of gyration (see Fig.~\ref{fig:ake_volume}), related to the configurational volume explored. Only the closed-state trajectory aligns with the fitted trend from the 11-protein dataset (dashed horizontal line in Fig.~\ref{fig:ake_volume}), consistent with the globular arrangement of the molecule in this state. By contrast, the full and open-state trajectories deviate from this reference value, yielding systematically lower $\nopt/\naa$.

These results indicate that $\nopt$ decreases as the average radius of gyration increases (see Fig.~\ref{fig:ake_volume}). In other words, conformational ensembles spanning larger configurational regions (e.g., the full trajectory or the open branch) require fewer atoms to be retained, whereas subsets confined to smaller fluctuations (\eg the closed branch) need more atoms to achieve the same information balance. This inverse relation is consistent with the notion, widely discussed in the context of dimensionality reduction and collective variable selection, that large-scale, highly collective motions can be described with a reduced number of degrees of freedom, while small-amplitude local fluctuations require higher resolution to be meaningfully distinguished \cite{amadei1993essential, hayward1998systematic, ichiye1991collective}.

In the specific case of AKE, previous work has shown that its hallmark open-to-closed transition is dominated by hinge-bending of the LID and NMP domains, a motion that is intrinsically low-dimensional \cite{henzler2007intrinsic}. Our findings quantitatively support this perspective: the broader the conformational exploration, the fewer atomic degrees of freedom are needed for an informative representation. In contrast, restricted conformational subsets entail finer local rearrangements, thus requiring higher resolution to preserve structural variability.

To further investigate the factors influencing the optimal resolution, we examined its relation to two intrinsic properties of the system: the distribution of atomic displacements, captured by the covariance matrix, and the information loss upon coarse-graining, quantified through the mapping entropy. These complementary analyses aim to disentangle how structural variability and the diversity of accessible conformations contribute to the emergence of a system-specific optimal resolution.

\subsection{Optimal number of sites \emph{vs.} position covariance}

\begin{figure*}
    \centering
    \includegraphics[width=\textwidth]{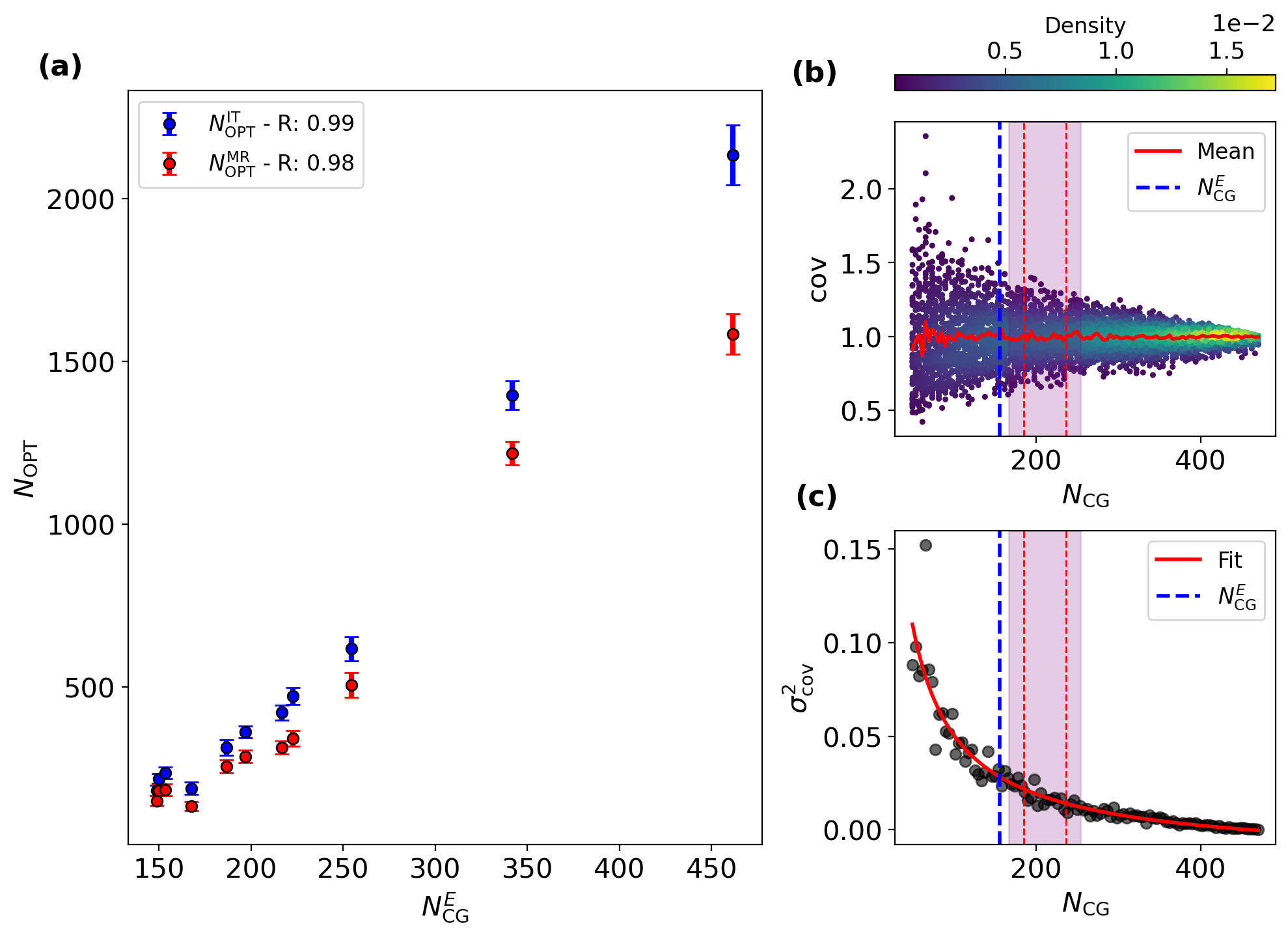}
  \caption{(a) Correlation between the elbow position (\( N_{\mathrm{CG}}^{E} \)) in the variance of the coarse-grained covariance (\( \sigma^2_{\mathrm{cov}} \)) and the optimal number of retained atoms \( N_{\mathrm{OPT}} \) identified by PROPRE, across a dataset of 11 proteins. Blue markers correspond to the information-theoretic estimate \( \nopt^{\mathrm{IT}} \) (slope \(\mu=-1\)), red markers to the relevance-maximum estimate \( \nopt^{\mathrm{MR}} \) (slope \(\mu=0\)). The legend reports the Pearson correlation coefficient for each criterion, and error bars indicate the standard deviation of \( N_{\mathrm{OPT}} \) within the most-probable bin returned by PROPRE. (b-c) Analysis of protein 1IGD. In both panels, vertical blue dashed line marks the elbow position (\( N_{\mathrm{CG}}^{E} \)), vertical dashed red lines delimit the core PROPRE interval, and the shaded region extends it by one standard deviation. (b) Distribution of the normalized coarse-grained covariance \( \mathrm{cov} \) over 50 random atom selections for each  $\ncg$ ; color encodes the pointwise density, the solid red line traces the running average. (c) Empirical variance \( \sigma^2_{\mathrm{cov}}(N_{\mathrm{CG}}) \) (gray dots) and fitted power-law decay (red curve).}
    \label{fig:cov_variance}
\end{figure*}

To evaluate how well a reduced representation captures the structural fluctuations of the full system, we analyzed the normalized trace of the covariance matrix, $\mathrm{cov}$, introduced in the Methods section. For each value of $\ncg$, \ie the number of retained atoms, $\mathrm{cov}$ was computed over 50 randomly-selected subsets of heavy atoms. The upper right panel of Fig.~\ref{fig:cov_variance} shows that this quantity fluctuates around 1, indicating that, on average, the magnitude of structural fluctuations is preserved even in reduced representations.

The spread of values around this mean decreases systematically with increasing $\ncg$, reflecting reduced variability among different atom selections as the representation becomes sufficiently detailed to capture most of the fluctuation profile. To quantify this trend, the variance $\sigma^2_{\mathrm{cov}}$ of the $\mathrm{cov}$ distribution  was computed for each value of $\ncg$. The lower right panel of Fig.~\ref{fig:cov_variance} reveals an elbow-shaped decay in $\sigma^2_{\mathrm{cov}}$, suggesting that beyond a certain resolution additional atoms contribute marginally to reproducing the system's fluctuations. The elbow position, $\ncg^\mathrm{E}$, was identified through an unsupervised procedure: the variance curve was fitted with a power-law function, and the elbow location, $\ncg^\mathrm{E}$,  was determined as the value of $\ncg$ that maximizes the distance between the fitted curve and the straight line connecting its endpoints. The shaded region in the same panel highlights the optimal resolution range identified by PROPRE, defined as the interval between the number of atoms maximizing relevance, $\noptRM$, and the point where the slope reaches -1, $\noptIT$. This interval consistently lies just beyond the elbow, suggesting that PROPRE tends to select slightly higher resolutions than those strictly required to retain the dominant structural fluctuations.

The plots in Fig.~\ref{fig:cov_variance} refer to a single representative protein, 1IGD, chosen for illustrative purposes to clarify the behaviour of the observables and the procedure.  This trend extends to the full dataset, as demonstrated by the scatter plot in the left panel, which compares the elbow positions with the optimal resolutions identified by PROPRE. A strong linear correlation is observed, with Pearson coefficients exceeding 0.98 for both the definition based on maximum-relevance ($\noptRM$) and -1 slope ($\noptIT$). These results confirm that the optimal resolution identified by PROPRE systematically lies just beyond the elbow position and scales linearly with the number of atoms required to adequately capture conformational variability.
Protein-specific analyses and detailed results for the entire dataset can be found in SI, Sec.IV.

\subsection{Optimal number of sites vs.\ mapping entropy}
As introduced in Methods section, the mapping entropy \( S_{\mathrm{map}} \) quantifies the loss of information incurred when projecting a molecular system onto a reduced representation. In this work, \( S_{\mathrm{map}} \) was computed for randomly-selected subsets of atoms to characterize the information loss associated with non-optimized CG mappings. For each value of \( \ncg \), 50 random subsets of heavy atoms were considered.

\begin{figure*}
    \centering
    \includegraphics[width=\textwidth]{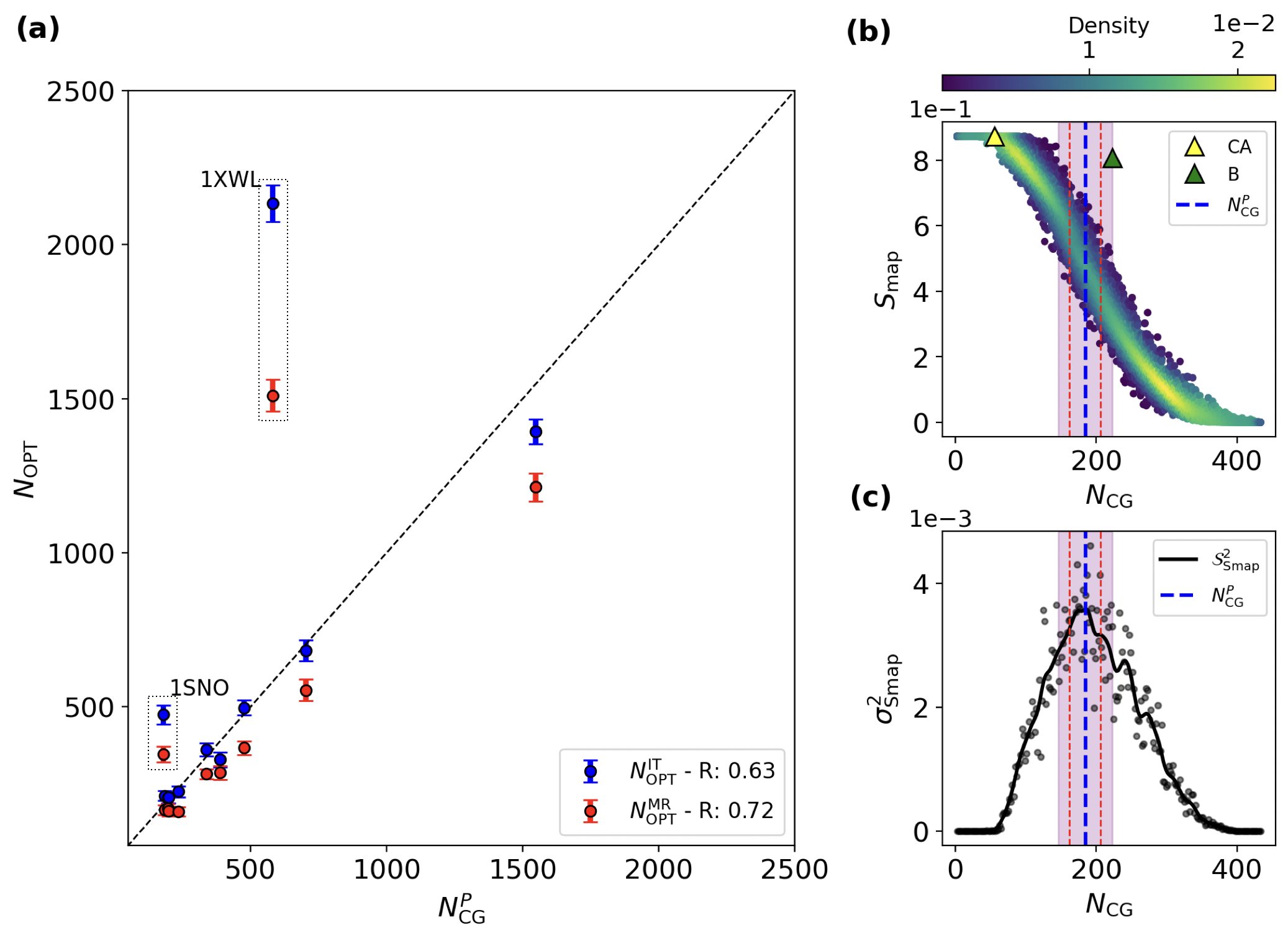}
    \caption{(a) Correlation between the peak position in the variance of the mapping entropy (\( N_{\mathrm{CG}}^{P} \)) and the optimal number of retained atoms \( N_{\mathrm{OPT}} \) identified by PROPRE, across a dataset of 11 proteins. Blue markers correspond to the information-theoretic estimate \( N_{\mathrm{OPT}}^{\mathrm{IT}} \) (slope \(\mu=-1\)), red markers to the relevance-maximum estimate \( N_{\mathrm{OPT}}^{\mathrm{MR}} \) (slope \(\mu=0\)). The legend reports the Pearson correlation coefficient for each criterion, and error bars indicate the standard deviation of \( N_{\mathrm{OPT}} \) within the most-probable bin returned by PROPRE. 
(b-c) Analysis of protein 1IGD. In both panels, the vertical blue dashed line marks the peak position \( N_{\mathrm{CG}}^{P} \), vertical dashed red lines delimit the core PROPRE interval, and the shaded region extends it by one standard deviation. 
(b) Distribution of the mapping entropy \( S_{\mathrm{map}} \) over 50 random atom selections for each \( N_{\mathrm{CG}} \); color encodes the pointwise density. The yellow and green triangles highlight the values of \( S_{\mathrm{map}} \) obtained from the specific selections of all C\(_\alpha\) atoms and all backbone atoms, respectively. 
(c) Empirical variance \( \sigma^2_{\mathrm{Smap}}(N_{\mathrm{CG}}) \) (gray dots) and its Gaussian-smoothed estimate \( \mathcal{S}^2_{\mathrm{Smap}} \) (black line).
}
    \label{fig:smap_variance}
\end{figure*}

The average value of \( S_{\mathrm{map}} \) decreases smoothly as \( \ncg \) increases, without exhibiting a distinct elbow or plateau. This trend indicates that, on average, adding more atoms progressively improves the reconstruction quality, but no sharp resolution threshold emerges from the mean behaviour. It is worth noting that when \( S_{\mathrm{map}} \) is computed for specific selections--namely all C\(_\alpha\) atoms or all backbone atoms---the resulting values lie well above those obtained from random heavy atom subsets of the same size (see Fig.~\ref{fig:smap_variance}.b, yellow triangle for C\(_\alpha\) and green triangle for backbone). This outcome indicates that such selections are not particularly informative for this dataset, as the proteins considered do not undergo large conformational rearrangements but instead sample local basins around their native globular structures, a type of exploration that is not well captured by C\(_\alpha\) or backbone coordinates alone.

By contrast, the variance of $S_{\mathrm{map}}$ across random selections--denoted $\sigma^2_{\mathrm{Smap}}$--exhibits a well-defined maximum. As shown in the lower-right panel of Fig.~\ref{fig:smap_variance}, $\sigma^2_{\mathrm{Smap}}$ is low at both extremes of $\ncg$ for opposite reasons: at very low $\ncg$, the number of retained atoms is too small to capture relevant information, so all mappings yield similarly high \( S_{\mathrm{map}} \) values; at high \( \ncg \), nearly all atoms are retained, and the mappings perform equivalently well, leading to low \( S_{\mathrm{map}} \). Between these limits, a distinct peak emerges, pointing to an intermediate resolution level at which information retention becomes highly sensitive to the specific atoms selected for the coarse-grained representation, so that alternative subsets of the same size can vary substantially in how well they capture the original system properties. In this regime, the large variance of $S_{\mathrm{map}}$ indicates that the amount of information preserved depends strongly on which atoms are chosen. When $\ncg$ is fixed well below $\nha$, targeted optimisation that minimises $S_{\mathrm{map}}$ by selecting which atoms to retain enjoys the greatest opportunity to reduce information loss and thereby enhance the coarse-grained representation without increasing resolution.

To robustly detect the location of this maximum, the raw variance profile \( \sigma^2_{\mathrm{Smap}} \) was smoothed using a Gaussian filter \cite{virtanen2020scipy} with a standard deviation of 5. The resulting curve, denoted \( \mathcal{S}^2_{\mathrm{Smap}} \), enables reliable identification of the peak resolution, labelled \( \ncg^P\).

The panels on the right of Fig.~\ref{fig:smap_variance} display the results for one representative protein (1IGD), showing both the variance profile and the resolution range identified by PROPRE. The vertical dashed red lines indicate the core PROPRE interval, defined between the number of atoms maximizing relevance and the point where the slope of the relevance-resolution curve equals \(-1\). The surrounding shaded band represents an extended interval that incorporates uncertainty--specifically, one standard deviation below the relevance maximum and above the slope \(-1\) point. Notably, the peak in \( \mathcal{S}^2_{\mathrm{Smap}} \) lies close to this shaded region, suggesting good consistency between the statistical properties of \( S_{\mathrm{map}} \) and the information-theoretic resolution inferred by PROPRE.  SI, Sec.V provides an extended presentation of the results for each individual protein in the dataset.

This relationship holds more generally across the entire dataset, as shown in the left panel of Fig.~\ref{fig:smap_variance}, which presents a scatter plot comparing the positions of largest $\sigma^2_{\mathrm{Smap}}$ (\( \ncg^{\mathrm{P}} \)) and the optimal number of sites \( \nopt \) determined by PROPRE for all 11 proteins. Despite two outliers, these quantities are positively correlated, with Pearson coefficients of $0.63$ and $0.72$ for the estimate based on \( \nopt^{\mathrm{IT}} \) (slope \(\mu=-1\)) and the one based on \( \nopt^{\mathrm{MR}} \) (largest relevance), respectively. When the two outliers (PDB IDs 1XWL and 1SNO) are excluded, the correlation between the variance peak position $\ncg^P$ and the PROPRE optimal resolution increases substantially, reaching a Pearson coefficient of $0.99$ for both estimates. This indicates that, for most cases, the optimal number of retained sites provides a reliable quantitative estimate of the position of largest \( S_{\mathrm{map}} \) variance.


\section{Conclusions}

In this work we have illustrated an algorithmic, unsupervised strategy that, taking the moves from general concepts of information theory, allows one to process a sample of protein configurations and identify the optimal fraction of heavy atoms to employ in a coarse-grained description of the system in order to have a sufficiently compact and yet informative picture of the molecule.

The resolution at which one inspects a complex system is not a neutral or irrelevant parameter. Whatever the level of detail employed to \emph{model} the system (quantum, classical, atomistic, coarse-grained...), the choice of how much of it to look at is inevitable (e.g. our brains cannot process all $3N$ coordinates of a large molecule at once) and fraught with consequences, as is any decision to keep something and discard something else. The selection of what to analyse, what to look at, what to preserve is subjected to the identification of the right \emph{amount} of detail to keep; in the case under examination here, it is the number of heavy atoms of a protein that it makes sense to retain in an optimal coarse-grained picture of the molecule.

The notion of \emph{optimality} employed here relies on the fact that a low-resolution representation -- specifically, a decimation mapping -- induces a clustering of the configurations, and that the statistical properties of such clustering depend on the number of atoms included in the mapping. By scanning through a range of retained atom numbers, one can measure the informativeness of the clustering and identify the most parsimonious yet informative resolution level.

We have applied this approach to a dataset of $11$ structurally and dynamically dissimilar proteins, as well as to adenylate kinase, an enzyme that interconverts between an open and a closed conformation. In general, we observed that the optimal number of retained atoms depends on system size, intrinsic flexibility, and the extent of conformational exploration.

More specifically, for the 11-protein dataset, a clear linear correlation emerges between system size and optimal resolution: the number of atoms to be retained increases proportionally with the number of residues. The slope of this correlation, close to four heavy atoms per amino acid, indicates the average amount of detail needed to preserve the essential information in the ensemble. The linearity of this relation is far from trivial: although it can be expected that larger proteins naturally require more atoms, there was no guarantee that there would have been a clear dependence, let alone a linear one. Moreover, this trend is particularly meaningful because it mirrors the average level of resolution adopted in widely used coarse-grained models, such as MARTINI \cite{souza2021martini} and SIRAH \cite{klein2023sirah}, thus offering a quantitative, data-driven basis for their empirical success. In MARTINI, four heavy atoms are typically mapped to a single coarse-grained bead, which for proteins translates to one backbone bead plus up to 4 side-chain beads per residue, depending on the size and polarity of the side chain. In SIRAH, a slightly finer resolution is employed, where 2 to 4 heavy atoms are typically mapped onto one interaction site. These intermediate levels of reduction have been shown to preserve the essential conformational space accessible to proteins, offering a good trade-off between computational speed and structural/thermodynamical accuracy over extended time and length scales. Importantly, a further reduction in resolution, below the four-to-one mapping, has been reported to compromise structural fidelity and the ability to capture key dynamical characteristics \cite{Noid2013Perspective, Kmiecik2016ChemRev}. For the proteins in the dataset, the optimal number of sites predicted by the PROPRE pipeline is intermediate between the number of CG beads obtained from the MARTINI and SIRAH mappings (see SI, sec III). Our finding that the data-driven fit identifies an effective scale similar to that of these CG models thus supports the view that $\sim4$ atoms per residue represents a natural balance point: it reduces complexity substantially while retaining the conformational ensemble characteristic of the atomistic system.

The case of adenylate kinase highlights that the optimal resolution is not dictated by size alone but also by the specific conformational arrangement of the molecule. When the trajectory is restricted to the closed basin, the scaling observed for the 11 proteins is recovered, consistent with the behaviour of a globular system undergoing limited rearrangements. In contrast, when the full trajectory or only the open branch is considered, the number of atoms required decreases, since large-scale collective motions can be described with fewer variables than the fine fluctuations within a single basin. This demonstrates that our investigation approach naturally accounts for the type of conformational variability at play, disentangling the contribution of system size from that of dynamics.

The results on the $11$ proteins were further validated through structural observables, confirming that the resolution levels identified preserve the configurational properties of the ensembles. The variance of the normalized covariance trace shows a characteristic elbow-like decay, with the optimal interval consistently located just beyond this point, where additional atoms provide only marginal improvements. A complementary confirmation comes from the mapping entropy, whose variance across random atom selections peaks at intermediate resolutions; the optimal interval falls precisely within the resolution range between highest relevance and slope $-1$, marking the regime where targeted optimization yields the greatest informational gain.
Taken together, these results demonstrate that the proposed strategy successfully identifies the minimal structural representation that preserves the essential configurational information contained in a protein ensemble. By highlighting the dependence on both protein size and conformational dynamics, it establishes a systematic criterion for defining resolution levels that are parsimonious yet maximally informative. In this way, it builds a quantitative bridge between atomistic simulations and coarse-grained models, offering an unsupervised, information-theoretic framework to guide multiscale representations of proteins.

The approach presented here is general and easy to employ; the outcome of its application to a sample of conformations of a protein does not only consist in a number of atoms to retain, but rather in a quantitative assessment of the amount of informative detail that one can extract out of this sample. In particular in the context of multiscale modelling, the identification of the smallest number of atoms that return a minimal yet informative picture of the molecule can serve the purpose of defining the natural resolution scale at which coarse-grained models should be built. In conclusion, the PROPRE approach represents a useful tool to shed new light on otherwise rather hidden features of a protein, helping researchers to build a quantitative, direct path between the properties of the system and the way one looks at it.

\begin{acknowledgement}

The authors are indebted with M. Giulini for an insightful reading of the manuscript and useful comments. 
RP acknowledges support from ICSC - Centro Nazionale di Ricerca in HPC, Big Data and Quantum Computing, funded by the European Union under NextGenerationEU. Views and opinions expressed are however those of the author(s) only and do not necessarily reflect those of the European Union or The European Research Executive Agency. Neither the European Union nor the granting authority can be held responsible for them.

\end{acknowledgement}

\begin{suppinfo}

Additional methodological details, full parameter lists, and supplementary analyses including relevance-resolution curve processing, PROPRE outputs, covariance and mapping entropy analyses, and comparison with coarse-grained models (PDF).

\end{suppinfo}

\section{Author contributions}
RP conceived the study and proposed the method. RF, GM and MM wrote the software. RF and TT ran the MD simulations. TT, GM, and MM performed the analysis. RP, TT, and MM interpreted the results. All authors drafted the paper, reviewed the results, and approved the final version of the manuscript.

\section{Data availability}
The raw data associated with this work are freely available on github at\\ \href{https://github.com/potestiolab/propre}{https://github.com/potestiolab/propre} and on the MaterialsCloud Archive at \\ \href{https://doi.org/10.24435/materialscloud:a7-r8}{https://doi.org/10.24435/materialscloud:a7-r8}.

\bibliography{achemso-demo}

\providecommand{\latin}[1]{#1}
\makeatletter
\providecommand{\doi}
  {\begingroup\let\do\@makeother\dospecials
  \catcode`\{=1 \catcode`\}=2 \doi@aux}
\providecommand{\doi@aux}[1]{\endgroup\texttt{#1}}
\makeatother
\providecommand*\mcitethebibliography{\thebibliography}
\csname @ifundefined\endcsname{endmcitethebibliography}
  {\let\endmcitethebibliography\endthebibliography}{}
\begin{mcitethebibliography}{61}
\providecommand*\natexlab[1]{#1}
\providecommand*\mciteSetBstSublistMode[1]{}
\providecommand*\mciteSetBstMaxWidthForm[2]{}
\providecommand*\mciteBstWouldAddEndPuncttrue
  {\def\EndOfBibitem{\unskip.}}
\providecommand*\mciteBstWouldAddEndPunctfalse
  {\let\EndOfBibitem\relax}
\providecommand*\mciteSetBstMidEndSepPunct[3]{}
\providecommand*\mciteSetBstSublistLabelBeginEnd[3]{}
\providecommand*\EndOfBibitem{}
\mciteSetBstSublistMode{f}
\mciteSetBstMaxWidthForm{subitem}{(\alph{mcitesubitemcount})}
\mciteSetBstSublistLabelBeginEnd
  {\mcitemaxwidthsubitemform\space}
  {\relax}
  {\relax}

\bibitem[Branden and Tooze(2012)Branden, and Tooze]{branden2012introduction}
Branden,~C.~I.; Tooze,~J. \emph{Introduction to protein structure}; Garland
  Science, 2012\relax
\mciteBstWouldAddEndPuncttrue
\mciteSetBstMidEndSepPunct{\mcitedefaultmidpunct}
{\mcitedefaultendpunct}{\mcitedefaultseppunct}\relax
\EndOfBibitem
\bibitem[Dill and MacCallum(2012)Dill, and MacCallum]{dill2012protein}
Dill,~K.~A.; MacCallum,~J.~L. \emph{Science} \textbf{2012}, \emph{338},
  1042--1046\relax
\mciteBstWouldAddEndPuncttrue
\mciteSetBstMidEndSepPunct{\mcitedefaultmidpunct}
{\mcitedefaultendpunct}{\mcitedefaultseppunct}\relax
\EndOfBibitem
\bibitem[Labeit and Kolmerer(1995)Labeit, and Kolmerer]{labeit1995titins}
Labeit,~S.; Kolmerer,~B. \emph{Science} \textbf{1995}, \emph{270},
  293--296\relax
\mciteBstWouldAddEndPuncttrue
\mciteSetBstMidEndSepPunct{\mcitedefaultmidpunct}
{\mcitedefaultendpunct}{\mcitedefaultseppunct}\relax
\EndOfBibitem
\bibitem[Leong \latin{et~al.}(2022)Leong, Lee, Mohtar, Syafruddin, Pung, and
  Low]{leong2022short}
Leong,~A. Z.-X.; Lee,~P.~Y.; Mohtar,~M.~A.; Syafruddin,~S.~E.; Pung,~Y.-F.;
  Low,~T.~Y. \emph{Journal of biomedical science} \textbf{2022}, \emph{29},
  19\relax
\mciteBstWouldAddEndPuncttrue
\mciteSetBstMidEndSepPunct{\mcitedefaultmidpunct}
{\mcitedefaultendpunct}{\mcitedefaultseppunct}\relax
\EndOfBibitem
\bibitem[Henzler-Wildman and Kern(2007)Henzler-Wildman, and
  Kern]{henzler2007dynamic}
Henzler-Wildman,~K.; Kern,~D. \emph{Nature} \textbf{2007}, \emph{450},
  964--972\relax
\mciteBstWouldAddEndPuncttrue
\mciteSetBstMidEndSepPunct{\mcitedefaultmidpunct}
{\mcitedefaultendpunct}{\mcitedefaultseppunct}\relax
\EndOfBibitem
\bibitem[Boehr \latin{et~al.}(2009)Boehr, Nussinov, and Wright]{boehr2009role}
Boehr,~D.~D.; Nussinov,~R.; Wright,~P.~E. \emph{Nature chemical biology}
  \textbf{2009}, \emph{5}, 789--796\relax
\mciteBstWouldAddEndPuncttrue
\mciteSetBstMidEndSepPunct{\mcitedefaultmidpunct}
{\mcitedefaultendpunct}{\mcitedefaultseppunct}\relax
\EndOfBibitem
\bibitem[Frauenfelder \latin{et~al.}(1991)Frauenfelder, Sligar, and
  Wolynes]{frauenfelder1991energy}
Frauenfelder,~H.; Sligar,~S.~G.; Wolynes,~P.~G. \emph{Science} \textbf{1991},
  \emph{254}, 1598--1603\relax
\mciteBstWouldAddEndPuncttrue
\mciteSetBstMidEndSepPunct{\mcitedefaultmidpunct}
{\mcitedefaultendpunct}{\mcitedefaultseppunct}\relax
\EndOfBibitem
\bibitem[Dill and Chan(1997)Dill, and Chan]{dill1997levinthal}
Dill,~K.~A.; Chan,~H.~S. \emph{Nature structural biology} \textbf{1997},
  \emph{4}, 10--19\relax
\mciteBstWouldAddEndPuncttrue
\mciteSetBstMidEndSepPunct{\mcitedefaultmidpunct}
{\mcitedefaultendpunct}{\mcitedefaultseppunct}\relax
\EndOfBibitem
\bibitem[Dill \latin{et~al.}(2008)Dill, Ozkan, Shell, and
  Weikl]{dill2008protein}
Dill,~K.~A.; Ozkan,~S.~B.; Shell,~M.~S.; Weikl,~T.~R. \emph{Annu. Rev.
  Biophys.} \textbf{2008}, \emph{37}, 289--316\relax
\mciteBstWouldAddEndPuncttrue
\mciteSetBstMidEndSepPunct{\mcitedefaultmidpunct}
{\mcitedefaultendpunct}{\mcitedefaultseppunct}\relax
\EndOfBibitem
\bibitem[McCammon \latin{et~al.}(1977)McCammon, Gelin, and
  Karplus]{mccammon1977dynamics}
McCammon,~J.~A.; Gelin,~B.~R.; Karplus,~M. \emph{nature} \textbf{1977},
  \emph{267}, 585--590\relax
\mciteBstWouldAddEndPuncttrue
\mciteSetBstMidEndSepPunct{\mcitedefaultmidpunct}
{\mcitedefaultendpunct}{\mcitedefaultseppunct}\relax
\EndOfBibitem
\bibitem[Karplus and McCammon(2002)Karplus, and McCammon]{karplus2002molecular}
Karplus,~M.; McCammon,~J.~A. \emph{Nature structural biology} \textbf{2002},
  \emph{9}, 646--652\relax
\mciteBstWouldAddEndPuncttrue
\mciteSetBstMidEndSepPunct{\mcitedefaultmidpunct}
{\mcitedefaultendpunct}{\mcitedefaultseppunct}\relax
\EndOfBibitem
\bibitem[Dror \latin{et~al.}(2012)Dror, Dirks, Grossman, Xu, and
  Shaw]{dror2012biomolecular}
Dror,~R.~O.; Dirks,~R.~M.; Grossman,~J.; Xu,~H.; Shaw,~D.~E. \emph{Annual
  review of biophysics} \textbf{2012}, \emph{41}, 429--452\relax
\mciteBstWouldAddEndPuncttrue
\mciteSetBstMidEndSepPunct{\mcitedefaultmidpunct}
{\mcitedefaultendpunct}{\mcitedefaultseppunct}\relax
\EndOfBibitem
\bibitem[Best and Hummer(2005)Best, and Hummer]{best2005reaction}
Best,~R.~B.; Hummer,~G. \emph{Proceedings of the National Academy of Sciences}
  \textbf{2005}, \emph{102}, 6732--6737\relax
\mciteBstWouldAddEndPuncttrue
\mciteSetBstMidEndSepPunct{\mcitedefaultmidpunct}
{\mcitedefaultendpunct}{\mcitedefaultseppunct}\relax
\EndOfBibitem
\bibitem[Peters and Trout(2006)Peters, and Trout]{peters2006obtaining}
Peters,~B.; Trout,~B.~L. \emph{The Journal of chemical physics} \textbf{2006},
  \emph{125}\relax
\mciteBstWouldAddEndPuncttrue
\mciteSetBstMidEndSepPunct{\mcitedefaultmidpunct}
{\mcitedefaultendpunct}{\mcitedefaultseppunct}\relax
\EndOfBibitem
\bibitem[No{\'e} and Clementi(2017)No{\'e}, and Clementi]{noe2017collective}
No{\'e},~F.; Clementi,~C. \emph{Current opinion in structural biology}
  \textbf{2017}, \emph{43}, 141--147\relax
\mciteBstWouldAddEndPuncttrue
\mciteSetBstMidEndSepPunct{\mcitedefaultmidpunct}
{\mcitedefaultendpunct}{\mcitedefaultseppunct}\relax
\EndOfBibitem
\bibitem[Chodera and No{\'e}(2014)Chodera, and No{\'e}]{chodera2014markov}
Chodera,~J.~D.; No{\'e},~F. \emph{Current opinion in structural biology}
  \textbf{2014}, \emph{25}, 135--144\relax
\mciteBstWouldAddEndPuncttrue
\mciteSetBstMidEndSepPunct{\mcitedefaultmidpunct}
{\mcitedefaultendpunct}{\mcitedefaultseppunct}\relax
\EndOfBibitem
\bibitem[No{\'e} and Fischer(2008)No{\'e}, and Fischer]{noe2008transition}
No{\'e},~F.; Fischer,~S. \emph{Current opinion in structural biology}
  \textbf{2008}, \emph{18}, 154--162\relax
\mciteBstWouldAddEndPuncttrue
\mciteSetBstMidEndSepPunct{\mcitedefaultmidpunct}
{\mcitedefaultendpunct}{\mcitedefaultseppunct}\relax
\EndOfBibitem
\bibitem[Mele \latin{et~al.}(2022)Mele, Covino, and
  Potestio]{mele2022information}
Mele,~M.; Covino,~R.; Potestio,~R. \emph{Soft Matter} \textbf{2022}, \emph{18},
  7064--7074\relax
\mciteBstWouldAddEndPuncttrue
\mciteSetBstMidEndSepPunct{\mcitedefaultmidpunct}
{\mcitedefaultendpunct}{\mcitedefaultseppunct}\relax
\EndOfBibitem
\bibitem[Noid(2013)]{Noid2013_PerspectiveCG}
Noid,~W.~G. \emph{The Journal of Chemical Physics} \textbf{2013}, \emph{139},
  090901\relax
\mciteBstWouldAddEndPuncttrue
\mciteSetBstMidEndSepPunct{\mcitedefaultmidpunct}
{\mcitedefaultendpunct}{\mcitedefaultseppunct}\relax
\EndOfBibitem
\bibitem[Saunders and Voth(2013)Saunders, and Voth]{saunders2013coarse}
Saunders,~M.~G.; Voth,~G.~A. \emph{Annual review of biophysics} \textbf{2013},
  \emph{42}, 73--93\relax
\mciteBstWouldAddEndPuncttrue
\mciteSetBstMidEndSepPunct{\mcitedefaultmidpunct}
{\mcitedefaultendpunct}{\mcitedefaultseppunct}\relax
\EndOfBibitem
\bibitem[Kmiecik \latin{et~al.}(2016)Kmiecik, Gront, Kolinski, Wieteska, Dawid,
  and Kolinski]{kmiecik2016coarse}
Kmiecik,~S.; Gront,~D.; Kolinski,~M.; Wieteska,~L.; Dawid,~A.~E.; Kolinski,~A.
  \emph{Chemical reviews} \textbf{2016}, \emph{116}, 7898--7936\relax
\mciteBstWouldAddEndPuncttrue
\mciteSetBstMidEndSepPunct{\mcitedefaultmidpunct}
{\mcitedefaultendpunct}{\mcitedefaultseppunct}\relax
\EndOfBibitem
\bibitem[Giulini \latin{et~al.}(2021)Giulini, Rigoli, Mattiotti, Menichetti,
  Tarenzi, Fiorentini, and Potestio]{giulini2021system}
Giulini,~M.; Rigoli,~M.; Mattiotti,~G.; Menichetti,~R.; Tarenzi,~T.;
  Fiorentini,~R.; Potestio,~R. \emph{Frontiers in Molecular Biosciences}
  \textbf{2021}, \emph{8}, 676976\relax
\mciteBstWouldAddEndPuncttrue
\mciteSetBstMidEndSepPunct{\mcitedefaultmidpunct}
{\mcitedefaultendpunct}{\mcitedefaultseppunct}\relax
\EndOfBibitem
\bibitem[Giulini \latin{et~al.}(2020)Giulini, Menichetti, Shell, and
  Potestio]{giulini2020information}
Giulini,~M.; Menichetti,~R.; Shell,~M.~S.; Potestio,~R. \emph{Journal of
  chemical theory and computation} \textbf{2020}, \emph{16}, 6795--6813\relax
\mciteBstWouldAddEndPuncttrue
\mciteSetBstMidEndSepPunct{\mcitedefaultmidpunct}
{\mcitedefaultendpunct}{\mcitedefaultseppunct}\relax
\EndOfBibitem
\bibitem[Chaimovich and Shell(2010)Chaimovich, and
  Shell]{ChaimovichShell2011_RelEntMapping}
Chaimovich,~A.; Shell,~M.~S. \emph{Physical Review E} \textbf{2010}, \emph{81},
  060104\relax
\mciteBstWouldAddEndPuncttrue
\mciteSetBstMidEndSepPunct{\mcitedefaultmidpunct}
{\mcitedefaultendpunct}{\mcitedefaultseppunct}\relax
\EndOfBibitem
\bibitem[Grigolon \latin{et~al.}(2016)Grigolon, Franz, and
  Marsili]{grigolon2016identifying}
Grigolon,~S.; Franz,~S.; Marsili,~M. \emph{Molecular BioSystems} \textbf{2016},
  \emph{12}, 2147--2158\relax
\mciteBstWouldAddEndPuncttrue
\mciteSetBstMidEndSepPunct{\mcitedefaultmidpunct}
{\mcitedefaultendpunct}{\mcitedefaultseppunct}\relax
\EndOfBibitem
\bibitem[Marsili and Roudi(2022)Marsili, and Roudi]{marsili2022quantifying}
Marsili,~M.; Roudi,~Y. \emph{Physics Reports} \textbf{2022}, \emph{963},
  1--43\relax
\mciteBstWouldAddEndPuncttrue
\mciteSetBstMidEndSepPunct{\mcitedefaultmidpunct}
{\mcitedefaultendpunct}{\mcitedefaultseppunct}\relax
\EndOfBibitem
\bibitem[Marsili \latin{et~al.}(2013)Marsili, Mastromatteo, and
  Roudi]{marsili2013sampling}
Marsili,~M.; Mastromatteo,~I.; Roudi,~Y. \emph{Journal of Statistical
  Mechanics: Theory and Experiment} \textbf{2013}, \emph{2013}, P09003\relax
\mciteBstWouldAddEndPuncttrue
\mciteSetBstMidEndSepPunct{\mcitedefaultmidpunct}
{\mcitedefaultendpunct}{\mcitedefaultseppunct}\relax
\EndOfBibitem
\bibitem[Cubero \latin{et~al.}(2019)Cubero, Jo, Marsili, Roudi, and
  Song]{cubero2019statistical}
Cubero,~R.~J.; Jo,~J.; Marsili,~M.; Roudi,~Y.; Song,~J. \emph{Journal of
  Statistical Mechanics: Theory and Experiment} \textbf{2019}, \emph{2019},
  063402\relax
\mciteBstWouldAddEndPuncttrue
\mciteSetBstMidEndSepPunct{\mcitedefaultmidpunct}
{\mcitedefaultendpunct}{\mcitedefaultseppunct}\relax
\EndOfBibitem
\bibitem[Cubero \latin{et~al.}(2020)Cubero, Marsili, and
  Roudi]{cubero2020multiscale}
Cubero,~R.~J.; Marsili,~M.; Roudi,~Y. \emph{Journal of computational
  neuroscience} \textbf{2020}, \emph{48}, 85--102\relax
\mciteBstWouldAddEndPuncttrue
\mciteSetBstMidEndSepPunct{\mcitedefaultmidpunct}
{\mcitedefaultendpunct}{\mcitedefaultseppunct}\relax
\EndOfBibitem
\bibitem[Holtzman \latin{et~al.}(2022)Holtzman, Giulini, and
  Potestio]{holtzman2022making}
Holtzman,~R.; Giulini,~M.; Potestio,~R. \emph{Physical Review E} \textbf{2022},
  \emph{106}, 044101\relax
\mciteBstWouldAddEndPuncttrue
\mciteSetBstMidEndSepPunct{\mcitedefaultmidpunct}
{\mcitedefaultendpunct}{\mcitedefaultseppunct}\relax
\EndOfBibitem
\bibitem[Sokal(1958)]{sokal1958statistical}
Sokal,~R.~R. \emph{Univ. Kansas, Sci. Bull.} \textbf{1958}, \emph{38},
  1409--1438\relax
\mciteBstWouldAddEndPuncttrue
\mciteSetBstMidEndSepPunct{\mcitedefaultmidpunct}
{\mcitedefaultendpunct}{\mcitedefaultseppunct}\relax
\EndOfBibitem
\bibitem[Shao \latin{et~al.}(2007)Shao, Tanner, Thompson, and
  Cheatham]{shao2007clustering}
Shao,~J.; Tanner,~S.~W.; Thompson,~N.; Cheatham,~T.~E. \emph{Journal of
  chemical theory and computation} \textbf{2007}, \emph{3}, 2312--2334\relax
\mciteBstWouldAddEndPuncttrue
\mciteSetBstMidEndSepPunct{\mcitedefaultmidpunct}
{\mcitedefaultendpunct}{\mcitedefaultseppunct}\relax
\EndOfBibitem
\bibitem[Maiorov and Crippen(1994)Maiorov, and
  Crippen]{maiorov1994significance}
Maiorov,~V.~N.; Crippen,~G.~M. \emph{Journal of molecular biology}
  \textbf{1994}, \emph{235}, 625--634\relax
\mciteBstWouldAddEndPuncttrue
\mciteSetBstMidEndSepPunct{\mcitedefaultmidpunct}
{\mcitedefaultendpunct}{\mcitedefaultseppunct}\relax
\EndOfBibitem
\bibitem[Kufareva and Abagyan(2012)Kufareva, and Abagyan]{kufareva2012methods}
Kufareva,~I.; Abagyan,~R. \emph{Homology Modeling: Methods and Protocols}
  \textbf{2012}, 231--257\relax
\mciteBstWouldAddEndPuncttrue
\mciteSetBstMidEndSepPunct{\mcitedefaultmidpunct}
{\mcitedefaultendpunct}{\mcitedefaultseppunct}\relax
\EndOfBibitem
\bibitem[Haimovici and Marsili(2015)Haimovici, and
  Marsili]{haimovici2015criticality}
Haimovici,~A.; Marsili,~M. \emph{Journal of Statistical Mechanics: Theory and
  Experiment} \textbf{2015}, \emph{2015}, P10013\relax
\mciteBstWouldAddEndPuncttrue
\mciteSetBstMidEndSepPunct{\mcitedefaultmidpunct}
{\mcitedefaultendpunct}{\mcitedefaultseppunct}\relax
\EndOfBibitem
\bibitem[Hensen \latin{et~al.}(2012)Hensen, Meyer, Haas, Rex, Vriend, and
  Grubm{\"u}ller]{hensen2012exploring}
Hensen,~U.; Meyer,~T.; Haas,~J.; Rex,~R.; Vriend,~G.; Grubm{\"u}ller,~H.
  \emph{PloS one} \textbf{2012}, \emph{7}, e33931\relax
\mciteBstWouldAddEndPuncttrue
\mciteSetBstMidEndSepPunct{\mcitedefaultmidpunct}
{\mcitedefaultendpunct}{\mcitedefaultseppunct}\relax
\EndOfBibitem
\bibitem[Schymkowitz \latin{et~al.}(2005)Schymkowitz, Borg, Stricher, Nys,
  Rousseau, and Serrano]{schymkowitz2005foldx}
Schymkowitz,~J.; Borg,~J.; Stricher,~F.; Nys,~R.; Rousseau,~F.; Serrano,~L.
  \emph{Nucleic acids research} \textbf{2005}, \emph{33}, W382--W388\relax
\mciteBstWouldAddEndPuncttrue
\mciteSetBstMidEndSepPunct{\mcitedefaultmidpunct}
{\mcitedefaultendpunct}{\mcitedefaultseppunct}\relax
\EndOfBibitem
\bibitem[Tarenzi \latin{et~al.}(2022)Tarenzi, Mattiotti, Rigoli, and
  Potestio]{tarenzi2022search}
Tarenzi,~T.; Mattiotti,~G.; Rigoli,~M.; Potestio,~R. \emph{Applied Sciences}
  \textbf{2022}, \emph{12}, 7157\relax
\mciteBstWouldAddEndPuncttrue
\mciteSetBstMidEndSepPunct{\mcitedefaultmidpunct}
{\mcitedefaultendpunct}{\mcitedefaultseppunct}\relax
\EndOfBibitem
\bibitem[Potestio \latin{et~al.}(2010)Potestio, Aleksiev, Pontiggia, Cozzini,
  and Micheletti]{potestio2010aladyn}
Potestio,~R.; Aleksiev,~T.; Pontiggia,~F.; Cozzini,~S.; Micheletti,~C.
  \emph{Nucleic acids research} \textbf{2010}, \emph{38}, W41--W45\relax
\mciteBstWouldAddEndPuncttrue
\mciteSetBstMidEndSepPunct{\mcitedefaultmidpunct}
{\mcitedefaultendpunct}{\mcitedefaultseppunct}\relax
\EndOfBibitem
\bibitem[David and Jacobs(2011)David, and Jacobs]{david2011characterizing}
David,~C.~C.; Jacobs,~D.~J. \emph{Journal of Molecular Graphics and Modelling}
  \textbf{2011}, \emph{31}, 41--56\relax
\mciteBstWouldAddEndPuncttrue
\mciteSetBstMidEndSepPunct{\mcitedefaultmidpunct}
{\mcitedefaultendpunct}{\mcitedefaultseppunct}\relax
\EndOfBibitem
\bibitem[Abraham \latin{et~al.}(2015)Abraham, Murtola, Schulz, P{\'a}ll, Smith,
  Hess, and Lindahl]{abraham2015gromacs}
Abraham,~M.~J.; Murtola,~T.; Schulz,~R.; P{\'a}ll,~S.; Smith,~J.~C.; Hess,~B.;
  Lindahl,~E. \emph{SoftwareX} \textbf{2015}, \emph{1}, 19--25\relax
\mciteBstWouldAddEndPuncttrue
\mciteSetBstMidEndSepPunct{\mcitedefaultmidpunct}
{\mcitedefaultendpunct}{\mcitedefaultseppunct}\relax
\EndOfBibitem
\bibitem[Lindorff-Larsen \latin{et~al.}(2010)Lindorff-Larsen, Piana, Palmo,
  Maragakis, Klepeis, Dror, and Shaw]{lindorff2010improved}
Lindorff-Larsen,~K.; Piana,~S.; Palmo,~K.; Maragakis,~P.; Klepeis,~J.~L.;
  Dror,~R.~O.; Shaw,~D.~E. \emph{Proteins: Structure, Function, and
  Bioinformatics} \textbf{2010}, \emph{78}, 1950--1958\relax
\mciteBstWouldAddEndPuncttrue
\mciteSetBstMidEndSepPunct{\mcitedefaultmidpunct}
{\mcitedefaultendpunct}{\mcitedefaultseppunct}\relax
\EndOfBibitem
\bibitem[Jorgensen \latin{et~al.}(1983)Jorgensen, Chandrasekhar, Madura, Impey,
  and Klein]{jorgensen1983comparison}
Jorgensen,~W.~L.; Chandrasekhar,~J.; Madura,~J.~D.; Impey,~R.~W.; Klein,~M.~L.
  \emph{The Journal of chemical physics} \textbf{1983}, \emph{79},
  926--935\relax
\mciteBstWouldAddEndPuncttrue
\mciteSetBstMidEndSepPunct{\mcitedefaultmidpunct}
{\mcitedefaultendpunct}{\mcitedefaultseppunct}\relax
\EndOfBibitem
\bibitem[Bussi \latin{et~al.}(2007)Bussi, Donadio, and
  Parrinello]{bussi2007canonical}
Bussi,~G.; Donadio,~D.; Parrinello,~M. \emph{The Journal of chemical physics}
  \textbf{2007}, \emph{126}, 014101\relax
\mciteBstWouldAddEndPuncttrue
\mciteSetBstMidEndSepPunct{\mcitedefaultmidpunct}
{\mcitedefaultendpunct}{\mcitedefaultseppunct}\relax
\EndOfBibitem
\bibitem[Parrinello and Rahman(1981)Parrinello, and
  Rahman]{parrinello1981polymorphic}
Parrinello,~M.; Rahman,~A. \emph{Journal of Applied physics} \textbf{1981},
  \emph{52}, 7182--7190\relax
\mciteBstWouldAddEndPuncttrue
\mciteSetBstMidEndSepPunct{\mcitedefaultmidpunct}
{\mcitedefaultendpunct}{\mcitedefaultseppunct}\relax
\EndOfBibitem
\bibitem[Hess \latin{et~al.}(1997)Hess, Bekker, Berendsen, and
  Fraaije]{hess1997lincs}
Hess,~B.; Bekker,~H.; Berendsen,~H.~J.; Fraaije,~J.~G. \emph{Journal of
  computational chemistry} \textbf{1997}, \emph{18}, 1463--1472\relax
\mciteBstWouldAddEndPuncttrue
\mciteSetBstMidEndSepPunct{\mcitedefaultmidpunct}
{\mcitedefaultendpunct}{\mcitedefaultseppunct}\relax
\EndOfBibitem
\bibitem[Humphrey \latin{et~al.}(1996)Humphrey, Dalke, and
  Schulten]{humphrey14vmd}
Humphrey,~W.; Dalke,~A.; Schulten,~K. \emph{J. Mol. Graph.} \textbf{1996},
  \emph{14}, 27--38\relax
\mciteBstWouldAddEndPuncttrue
\mciteSetBstMidEndSepPunct{\mcitedefaultmidpunct}
{\mcitedefaultendpunct}{\mcitedefaultseppunct}\relax
\EndOfBibitem
\bibitem[Shell(2008)]{shell2008relative}
Shell,~M.~S. \emph{The Journal of chemical physics} \textbf{2008}, \emph{129},
  144108\relax
\mciteBstWouldAddEndPuncttrue
\mciteSetBstMidEndSepPunct{\mcitedefaultmidpunct}
{\mcitedefaultendpunct}{\mcitedefaultseppunct}\relax
\EndOfBibitem
\bibitem[Aldrigo \latin{et~al.}(2025)Aldrigo, Menichetti, and
  Potestio]{aldrigo2025low}
Aldrigo,~R.; Menichetti,~R.; Potestio,~R. \emph{Physical Review E}
  \textbf{2025}, \emph{111}, 044315\relax
\mciteBstWouldAddEndPuncttrue
\mciteSetBstMidEndSepPunct{\mcitedefaultmidpunct}
{\mcitedefaultendpunct}{\mcitedefaultseppunct}\relax
\EndOfBibitem
\bibitem[Giulini \latin{et~al.}(2024)Giulini, Fiorentini, Tubiana, Potestio,
  and Menichetti]{giulini2024excogito}
Giulini,~M.; Fiorentini,~R.; Tubiana,~L.; Potestio,~R.; Menichetti,~R.
  \emph{Journal of Chemical Information and Modeling} \textbf{2024}, \emph{64},
  4912--4927\relax
\mciteBstWouldAddEndPuncttrue
\mciteSetBstMidEndSepPunct{\mcitedefaultmidpunct}
{\mcitedefaultendpunct}{\mcitedefaultseppunct}\relax
\EndOfBibitem
\bibitem[Chaimovich and Shell(2011)Chaimovich, and Shell]{chaimovich2011coarse}
Chaimovich,~A.; Shell,~M.~S. \emph{The Journal of chemical physics}
  \textbf{2011}, \emph{134}\relax
\mciteBstWouldAddEndPuncttrue
\mciteSetBstMidEndSepPunct{\mcitedefaultmidpunct}
{\mcitedefaultendpunct}{\mcitedefaultseppunct}\relax
\EndOfBibitem
\bibitem[Amadei \latin{et~al.}(1993)Amadei, Linssen, and
  Berendsen]{amadei1993essential}
Amadei,~A.; Linssen,~A.~B.; Berendsen,~H.~J. \emph{Proteins: Structure,
  Function, and Bioinformatics} \textbf{1993}, \emph{17}, 412--425\relax
\mciteBstWouldAddEndPuncttrue
\mciteSetBstMidEndSepPunct{\mcitedefaultmidpunct}
{\mcitedefaultendpunct}{\mcitedefaultseppunct}\relax
\EndOfBibitem
\bibitem[Hayward and Berendsen(1998)Hayward, and
  Berendsen]{hayward1998systematic}
Hayward,~S.; Berendsen,~H.~J. \emph{Proteins: structure, function, and
  bioinformatics} \textbf{1998}, \emph{30}, 144--154\relax
\mciteBstWouldAddEndPuncttrue
\mciteSetBstMidEndSepPunct{\mcitedefaultmidpunct}
{\mcitedefaultendpunct}{\mcitedefaultseppunct}\relax
\EndOfBibitem
\bibitem[Ichiye and Karplus(1991)Ichiye, and Karplus]{ichiye1991collective}
Ichiye,~T.; Karplus,~M. \emph{Proteins: Structure, Function, and
  Bioinformatics} \textbf{1991}, \emph{11}, 205--217\relax
\mciteBstWouldAddEndPuncttrue
\mciteSetBstMidEndSepPunct{\mcitedefaultmidpunct}
{\mcitedefaultendpunct}{\mcitedefaultseppunct}\relax
\EndOfBibitem
\bibitem[Henzler-Wildman \latin{et~al.}(2007)Henzler-Wildman, Thai, Lei, Ott,
  Wolf-Watz, Fenn, Pozharski, Wilson, Petsko, and
  Karplus]{henzler2007intrinsic}
Henzler-Wildman,~K.~A.; Thai,~V.; Lei,~M.; Ott,~M.; Wolf-Watz,~M.; Fenn,~T.;
  Pozharski,~E.; Wilson,~M.~A.; Petsko,~G.~A.; Karplus,~M. \emph{Nature}
  \textbf{2007}, \emph{450}, 838--844\relax
\mciteBstWouldAddEndPuncttrue
\mciteSetBstMidEndSepPunct{\mcitedefaultmidpunct}
{\mcitedefaultendpunct}{\mcitedefaultseppunct}\relax
\EndOfBibitem
\bibitem[Virtanen \latin{et~al.}(2020)Virtanen, Gommers, Oliphant, Haberland,
  Reddy, Cournapeau, Burovski, Peterson, Weckesser, and
  Bright]{virtanen2020scipy}
Virtanen,~P.; Gommers,~R.; Oliphant,~T.~E.; Haberland,~M.; Reddy,~T.;
  Cournapeau,~D.; Burovski,~E.; Peterson,~P.; Weckesser,~W.; Bright,~J.
  \emph{Nature methods} \textbf{2020}, \emph{17}, 261--272\relax
\mciteBstWouldAddEndPuncttrue
\mciteSetBstMidEndSepPunct{\mcitedefaultmidpunct}
{\mcitedefaultendpunct}{\mcitedefaultseppunct}\relax
\EndOfBibitem
\bibitem[Souza \latin{et~al.}(2021)Souza, Alessandri, Barnoud, Thallmair,
  Faustino, Gr{\"u}newald, Patmanidis, Abdizadeh, Bruininks, and
  Wassenaar]{souza2021martini}
Souza,~P.~C.; Alessandri,~R.; Barnoud,~J.; Thallmair,~S.; Faustino,~I.;
  Gr{\"u}newald,~F.; Patmanidis,~I.; Abdizadeh,~H.; Bruininks,~B.~M.;
  Wassenaar,~T.~A. \emph{Nature methods} \textbf{2021}, \emph{18},
  382--388\relax
\mciteBstWouldAddEndPuncttrue
\mciteSetBstMidEndSepPunct{\mcitedefaultmidpunct}
{\mcitedefaultendpunct}{\mcitedefaultseppunct}\relax
\EndOfBibitem
\bibitem[Klein \latin{et~al.}(2023)Klein, So{\~n}ora, Santos, Frigini,
  Ballesteros-Casallas, Machado, and Pantano]{klein2023sirah}
Klein,~F.; So{\~n}ora,~M.; Santos,~L.~H.; Frigini,~E.~N.;
  Ballesteros-Casallas,~A.; Machado,~M.~R.; Pantano,~S. \emph{Journal of
  structural biology} \textbf{2023}, \emph{215}, 107985\relax
\mciteBstWouldAddEndPuncttrue
\mciteSetBstMidEndSepPunct{\mcitedefaultmidpunct}
{\mcitedefaultendpunct}{\mcitedefaultseppunct}\relax
\EndOfBibitem
\bibitem[Noid(2013)]{Noid2013Perspective}
Noid,~W.~G. \emph{The Journal of Chemical Physics} \textbf{2013}, \emph{139},
  090901\relax
\mciteBstWouldAddEndPuncttrue
\mciteSetBstMidEndSepPunct{\mcitedefaultmidpunct}
{\mcitedefaultendpunct}{\mcitedefaultseppunct}\relax
\EndOfBibitem
\bibitem[Kmiecik \latin{et~al.}(2016)Kmiecik, Gront, Kolinski, Wieteska, Dawid,
  and Kolinski]{Kmiecik2016ChemRev}
Kmiecik,~S.; Gront,~D.; Kolinski,~M.; Wieteska,~L.; Dawid,~A.~E.; Kolinski,~A.
  \emph{Chemical Reviews} \textbf{2016}, \emph{116}, 7898--7936\relax
\mciteBstWouldAddEndPuncttrue
\mciteSetBstMidEndSepPunct{\mcitedefaultmidpunct}
{\mcitedefaultendpunct}{\mcitedefaultseppunct}\relax
\EndOfBibitem
\end{mcitethebibliography}

\end{document}